\documentclass[pra,aps,twocolumn,reprint,floatfix,showpacs,10pt]{revtex4-1}
\usepackage{graphicx}
\usepackage{amsmath}
\usepackage[colorlinks,bookmarks=false,citecolor=blue,linkcolor=red,urlcolor=black]{hyperref}
\usepackage{umoline}

\begin{document}
\title{Two bosonic quantum walkers in one-dimensional optical lattices}
\author{Dariusz Wiater$^1$}
\email{dariusz.wiater@uj.edu.pl}
\author{Tomasz Sowi\'nski$^2$}
\email{tomasz.sowinski@ifpan.edu.pl}
\author{Jakub Zakrzewski$^{1,3}$} 
 \email{kuba@if.uj.edu.pl}

\affiliation{
\mbox{$^1$ Instytut Fizyki imienia Mariana Smoluchowskiego, Uniwersytet Jagiello{\'n}ski, ulica \L{}ojasiewicza 11, PL-30059 Krak\'ow, Poland }
\mbox{$^2$ Institute of Physics, Polish Academy of Sciences, Aleja Lotnikow 32/46, PL-02668 Warsaw, Poland}
\mbox{$^3$ Mark Kac Complex Systems Research Center,
Uniwersytet Jagiello\'nski, Krak\'ow, Poland }
}
\begin{abstract}
Dynamical properties of two bosonic quantum walkers in a one-dimensional lattice are studied theoretically. Depending on the initial state, interactions, lattice tilting, and lattice disorder, whole plethora of different behaviors are observed. Particularly, it is shown that two bosons system manifests the many-body localization like behavior in the presence of a quenched disorder. The whole analysis is based on a specific decomposition of the temporal density profile into different contributions from singly and doubly occupied sites. In this way, the role of interactions is extracted. Since the contributions can be directly measured in experiments with ultra-cold atoms in optical lattices, the predictions presented may have some importance for upcoming experiment.
\end{abstract}
\date{\today}

\maketitle

\section{Introduction}
Quantum walks are quantum analog of classical random walks. Usually, they are split into discrete and continuous walks. Discrete cases refer to situations when one considers quantum version of a coin flipping after each step \cite{QuantumWalks}. In contrast, continuous quantum walks have a different character  based on an appropriate time evolution equation. For quantum systems the evolution is determined by the  Schr\"{o}dinger equation and together with the Hamiltonian they govern behavior of quantum walkers. There are different possibilities related to physical implementation of quantum walks, especially connected with quantum optics experiments \cite{Mendez93}. One of such realizations takes place in optical lattices -- atomic physics systems which imitate structures known from condensed matter physics. Periodic optical potentials are obtained by appropriate standing-wave configurations of laser beams. Huge advantage of such systems relays on a fact that many parameters of optical lattices can be tuned 
and controlled with a very high accuracy. For that reason atoms in optical lattice may serve as a powerful tool to simulate phenomena from different branches of physics \cite{Lewenstein07,Lewenstein12}. 
{Here the expansion of interacting particles from well controlled initial states brings information about the nature of many-body dynamics (see e.g. \cite{Ronzheimer2013,Song2014,Vidmar2015,Hauschild2015}.}

{Few particles quantum walks have been quite intensively studied both for photonic and spin systems \cite{Peruzzo2010,Bromberg2009,Lahini2010,Lahini2012,Solntsev2012,Benedetti2012,Fukuhara2013,Ganahl2012,Meinecke2013,Liu2014}.} 
Recent beautiful experiments \cite{Preiss2015} have demonstrated a high controllability of quantum walks of atoms in optical lattices and nice agreement between theoretical simulations and experimental results. One may consider both fermionic and bosonic quantum walks. For strongly interacting bosons, close to the Tonks-Girardeau regime, one may observe effective fermionization of the bosonic motion. In the pure fermionic case walks accompanied by spin-flipping were also analyzed \cite{Wang2015}. 
{The role of interactions and statistics has been considered in \cite{Qin2014}. Recent studies expand the quantum walk studies also to noisy (time dependent) systems \cite{Benedetti2016,Siloi2017,Piccinini2017}.}

In this paper we come back to the problem of quantum bosonic walkers in one-dimensional optical lattices. In the standard approach they are described by the old-fashioned Bose-Hubbard hamiltonian \cite{Jaksch1998,Zwerger2003,Dutta2015}, 
\begin{equation}\label{hamBH}
\hat H_\mathtt{BH}= -J \sum_{i=-L}^{L-1} \left(\hat a_{i+1}^{\dagger} \hat a_i +h.c.\right)+ \frac{U}{2} \sum_{i=-L}^L  \hat n_i(\hat n_i -1),
\end{equation}
where $\hat a_i$ is a bosonic operator annihilating  particle at site $i$ and $\hat n_i=\hat a_i^{\dagger}\hat a_i$ is a {particle number} (density) operator. The first term {describes} tunneling between neighboring sites and the second {term the} on-site interactions. Without loosing generality, we assume $J=1$ which fixes the energy (time) unit. 
Since the Hamiltonian \eqref{hamBH} commutes with the total number of particles $\hat N = \sum_i \hat n_i$  {operator}, the analysis can be performed independently in subspaces of a given number of bosons.  In the following we shall concentrate on the influence of interactions on quantum walkers.  To make our study as comprehensive as possible we consider different initial quantum states $|\mathtt{ini}\rangle$ and different  perturbations of the model. We assume a  family of initial states, mainly the states in which particles form a gaussian beams centered around a chosen site of the lattice, $k_0$. Depending of the number of particles in the system $N$ the family is defined as
\begin{equation} \label{IniState}
|\mathtt{ini}_\sigma\rangle = {\cal N}\left(\sum_k\mathrm{e}^{-(k-k_0)^2/2\sigma}\,\hat{a}_k^\dagger\right)^N |\mathtt{vac}\rangle,
\end{equation}
where ${\cal N}$ is a normalization constant, $k_0$ and $\sigma$ are the position of a center and the spread of the gaussian state, respectively.  Note that in the limiting situation, $\sigma\rightarrow 0$, the state with a single site occupied by all particles $(\hat a_0^\dagger)^N|\mathtt{vac}\rangle$ is obtained. In the case of two bosons, we also consider a situation that bosons initially occupy adjacent sites
\begin{equation} \label{IniState2}
|\mathtt{ini}'\rangle = \hat{a}_{k_0+1}^\dagger \hat{a}_{k_0}^\dagger |\mathtt{vac}\rangle.
\end{equation}
Typically we start the evolution from the center of the lattice i.e $k_0=0$.

As external perturbations of the model we take into account three different effects which can be described by the sum of three following terms:
\begin{equation} \label{Ham}
\hat{H}_\mathtt{ext} =  \hat{T} + \hat{V} + \hat{D},
\end{equation}
where
\begin{subequations} \label{Ham2}
\begin{align}
\hat{T} &= F \sum_i i\, \hat a_i^{\dagger}\,  \hat a_i, \label{Ham2a}\\
\hat{V} &= V\sum_{i}\sum_{k\ne 0} k^{-\alpha}\, \hat n_i\hat n_{i+k}, \label{Ham2b}\\
\hat{D} &= \lambda \sum_{i} \cos\left[2 \pi (\tau i +\phi)\right] \hat a_i^{\dagger}  \hat a_i. \label{Ham2c}
\end{align}
\end{subequations}
$\hat T$ describes linear tilt of the optical lattice. It mimics an existence of a uniform external electric field in the system. The second term $\hat V$ takes into account the most relevant contribution from long-range interactions. Depending on parameter $\alpha$ different long-range potentials are described. For example, $\alpha=1$ corresponds to Coulomb-like behavior, whereas $\alpha=3$ is typical for dipole-dipole interactions. It is worth noticing that in the case of long-range interactions it may be necessary to take into account other additional terms related to density dependent tunnelings \cite{Dutta2015}. Those however, will not be relevant for simple quantum walkers we discuss here. The third term $\hat D$ is introduced to mimics random disorder in the system. Its form is explained and its influence on  the dynamical properties of the system is described in Sec. \ref{AN}.

One should keep in mind that the physics of interacting bosons in tilted optical lattices have been extensively studied in the past (some representative references include \cite{Gluck2002,Sachdev2002,Kolovsky2003,Buchleitner2003,Kolovsky04,Kolovsky2010,Diaz2013,Mandt2014,Meinert2014}). 

The paper is organized as follows. In Section~\ref{BH} we shortly discuss our numerical approach to the problem and we introduce the concept of partial density contributions describing single and double occupations noting that they can be measured directly in experiments. In Section~\ref{Results} we analyze different dynamical properties of two interacting walkers depending on their initial state.  The analysis is performed for different arrangements of the lattice and different strengths and types of interactions repeating, and in some cases expanding  the results of \cite{Preiss2015}. In particular, we show that particular components to the density profile behave differently in the presence of interactions which can be used as good indicator of {the role of interactions}. In Section~\ref{AN} we discuss the effect of disorder on quantum walkers -- the subject, as far as we know, not analyzed before. We show that quantum walks even for two particles only, reveal similar characteristics as many-body 
localization \cite{Huse14,Rahul15}. This allows us to claim that the approach presented is an interesting  way to observe quantum dynamics in disordered cases. We show a strong two-body localization in the case of tilted lattices. We conclude in Section\ref{Concl}.

\section{The Method}\label{BH}  
Numerical simulations are performed using the numerically exact diagonalization of the many-body Hamiltonian in the Fock  subspace of a given number of particles. 
 The time evolution is expressed in terms of  eigenstates $|\psi_m\rangle$ and the corres ponding eigenenergies ${\cal E}_m$ in a textbook manner:
\begin{equation}
|\psi(t)\rangle = \sum_m \langle\psi_m|\mathtt{ini}_\sigma\rangle\,\mathrm{e}^{-i {\cal E}_m t}\, |\psi_m\rangle.
\end{equation}
{During time evolution the wavepacket, initially localized in the center of our system spreads and eventually could reach the borders at $\pm L$ spoiling the numerical results.
We terminate the time evolution well before reaching the borders.}

The main quantity which we focus on is the density distribution of bosons among lattice sites. It can be calculated directly from a temporal state of the system 
\begin{equation}
\label{ni}
 n(i)=\langle\psi(t)|\hat a_i^{\dagger}  \hat a_i | \psi(t)\rangle.
\end{equation}
To  better understand the behavior of the system we view this quantity as a hierarchal sum of densities of different local occupations, i.e., $n(i) = n_1(i) + n_2(i) + n_3(i) +\ldots$, where
consecutive densities $n_k(i)$ are calculated according to \eqref{ni} provided that local occupation of $i$-th site is exactly equal to $k$. In the case of single-boson problem there is only one contribution to the density $n(i) = n_1(i)$. For the case of two bosons one {may} represent the density as a sum of two contributions $n(i) = n_1(i) + n_2(i)$ where
\begin{subequations}
\begin{align}\label{n2}
n_2(i)&=\langle\psi(t)|\hat a_i^{\dagger}\hat a_i^{\dagger}  \hat a_i  \hat a_i| \psi(t)\rangle, \\
n_1(i)&=n(i)-n_2(i).
\end{align}
\end{subequations}
Generalizations to a larger number of particles is straightforward.

With this specific decomposition of the density distribution we are able to study an influence of inter-particle interactions to the behavior of the system. Since different contributions $n_k(i)$ have different sensitivity to the strength of on-site interactions their behavior is a quite good indicator of a role of interactions in the system.
\begin{figure}
\includegraphics[width=\columnwidth]{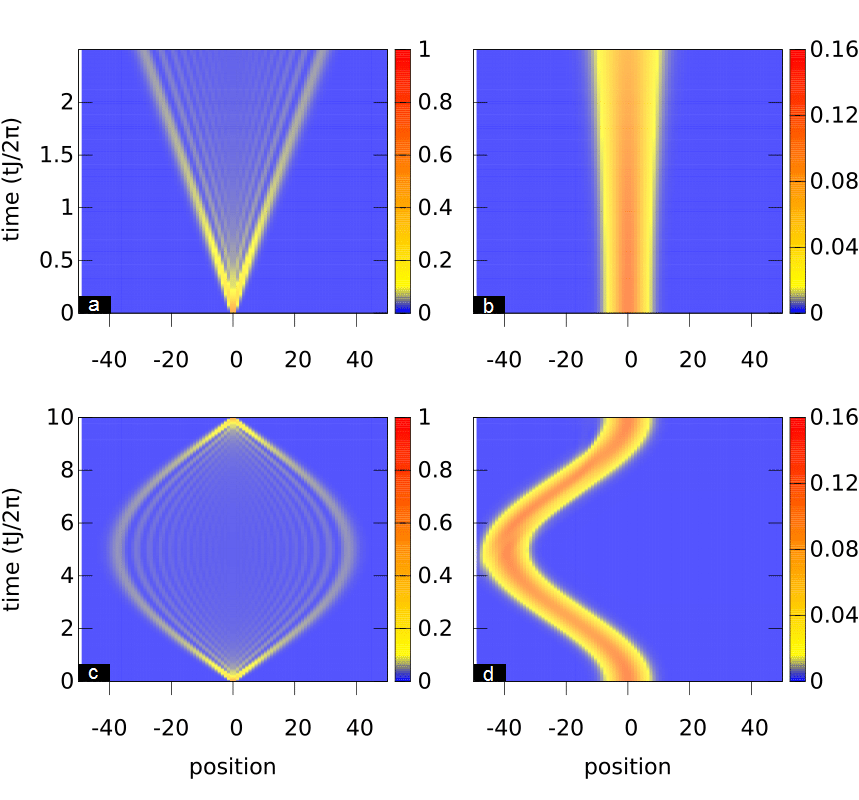}
\caption{Time evolution of the density distribution $n(i)$ in the case of a single particle quantum walk. Left panel: initially the particle occupies  single lattice site. Right panel: initially the particle is in a gaussian state $|\mathtt{ini}_\sigma\rangle$ with $\sigma^2=25$. Depending on a lattice tilting one observes ballistic expansion across the lattice (upper panel for $F=0$) or characteristic Bloch oscillations (bottom panel for $F=0.5$). Note, that depending on the initial state, Bloch oscillations are related to different evolution of the center of the density distribution.}
 \label{Fig1}
\end{figure}

\section{The results} \label{Results}
In this section we describe dynamical properties of the system in a complete absence of the disorder, i.e. we assume that $\hat{D}\equiv 0$. The role of the disorder and its impact to different properties of the system studied are shown in Sec. \ref{AN}.

{\bf Single-particle diffusion.--}
First let us remind well known results for the simplest possible quantum walk, i.e., ballistic dynamics of a single quantum particle in a periodic potential. Whenever system is described by the simplest Bose-Hubbard Hamiltonian \eqref{hamBH} the initially localized wave function of the particle is spreading across a whole lattice. Time evolution of the density distribution $n(i)$ depends however crucially on initial state of the particle. As shown in Fig.~\ref{Fig1}: {\it (i)} when particle is initially isolated in a chosen lattice site then the dynamics is strongly affected by the presence of the lattice shape and characterized interference pattern is visible during an expansion; {\it (ii)} when particle is essentially delocalized, i.e., characteristic width of the wave function is much larger than the distance between lattice sites, the dynamics is described almost perfectly by continuous counterpart of the Schr\"odinger equation. In the former case a characteristic spreading of the gaussian state is 
visible.

It is worth noticing that, in the case of single-particle ballistic expansion, the time evolution of the density distribution is described analytically by the Bessel function of the first kind 
$n_i={\cal J}^2_i(2Jt)$ \cite{Hartmann2004}  with the speed of an expansion characterized by $J$.

The situation is distinctly different in the presence of additional lattice tilting $\hat T$ described by the Hamiltonian \eqref{Ham2a}. In this case, characteristic oscillatory evolution of the density distribution is present. This counter-intuitive behavior, known as Bloch oscillations, is a direct consequence of the band structure of a periodic potential \cite{Gluck2002,Dias07,Khomeriki10}. Therefore, it is a generic behavior in any periodic structure affected by a constant external force. Depending on the initial state of the system, Bloch oscillations are manifested in various ways. If the initial state is localized, in the sense that its natural spatial width is of the order of the lattice length, then oscillations are left-right symmetric, i.e. the density profile alternately expands and shrinks around the initial position (see Fig. \ref{Fig1}c). 
On the other hand, whenever the initial state is wide enough, an oscillatory behavior is associated with the position of the center of the wave-packet. During a whole evolution the width of the packet is preserved (see Fig. \ref{Fig1}d). A detailed discussion of that phenomenon is presented in \cite{Hartmann2004}.

Periodic in time Bloch oscillations are especially interesting for potential applications taking 
into account that the experimental control over such 
oscillations has been achieved successfully \cite{Mendez93, Dahan96,Lyssenko97,Zanesini09,Poli11}.

{\bf Dynamics of two interacting bosons.--}
The physics of the system for larger number of particles is, of course, much richer. Depending on the quantum statistics and mutual interactions assumed dynamical properties of the system can be completely different. Here, we concentrate on the simplest extension of the single-particle model, i.e., the model with two interacting spinless bosons. Again we analyze the impact of the external field $F$ but an additional parameter affecting the picture is the interaction strength $U$. Note also, that in the case of two particles, there is much more freedom in defining an initial state. 
\begin{figure}
\includegraphics[width=\columnwidth]{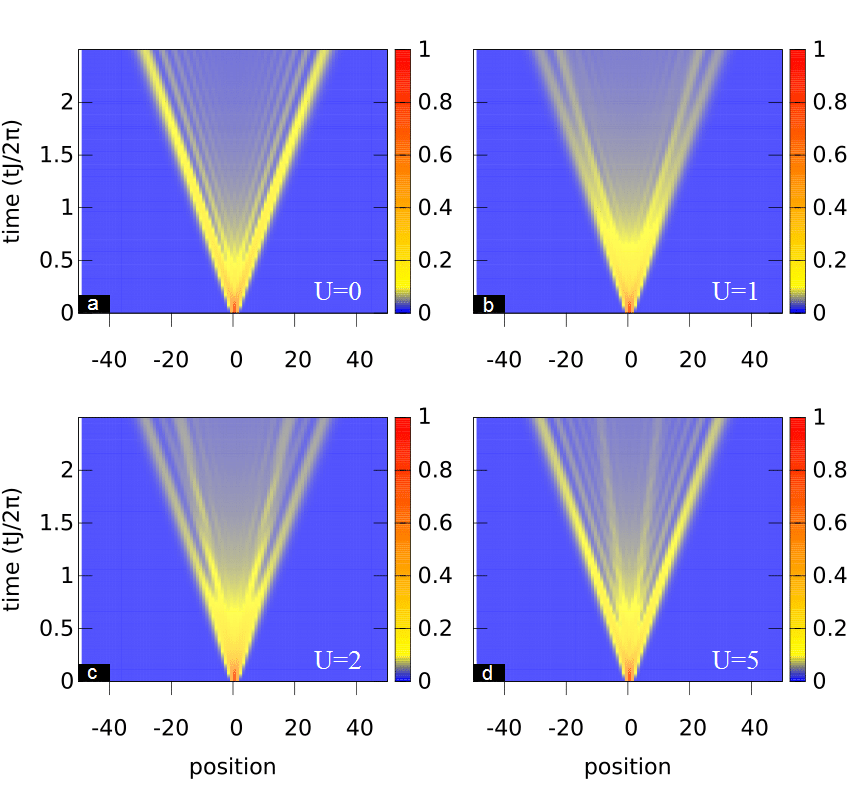}
\caption{Time evolution of the density distribution $n(i)$ for two interacting bosonic walkers in the case of vanishing lattice tilting (for $F=0$). Depending on strength of the on-site interaction, (a) the single-particle diffusion is restored or (b)-(d) additional inner density cone is visible. Note, that the apex angle of the inner cone, in contrast to outer cone, strongly depends on interactions. This fact is reflected in a different behaviors of densities of singly and doubly occupied sites (see Fig.~\ref{Fig3}).}
\label{Fig2}
\end{figure}

To give first insight to the two-particle problem let us first focus on the situation in which bosons occupy adjacent sites, i.e., the initial state is given by \eqref{IniState2}. In this case, the initial state has always the same energy independently on strength of on-site interactions $U$, therefore comparison to the noninteracting case is simplified.

In Fig. \ref{Fig2} we show an evolution of the density distribution in the absence of external force $F$ and different interactions $U$. As it is seen, in the absence of interactions (Fig. \ref{Fig2}a) the resulting evolution is fully consistent with the single-particle case (compare with Fig.\ref{Fig1}a). Both particles independently spread across the lattice. Whenever interactions are switched on, then a specific fragmentarization of the distribution flow is observed (Fig. \ref{Fig2}b-d). It is visible that for stronger interactions an evolution of the density profile $n(i)$ reveals a new component with slower spread in time. To make better understanding of this phenomenon, we analyze independent components of the density distribution originating in singly and doubly occupied sites, $n_1(i)$ and $n_2(i)$, respectively.  
\begin{figure}
\includegraphics[width=\columnwidth]{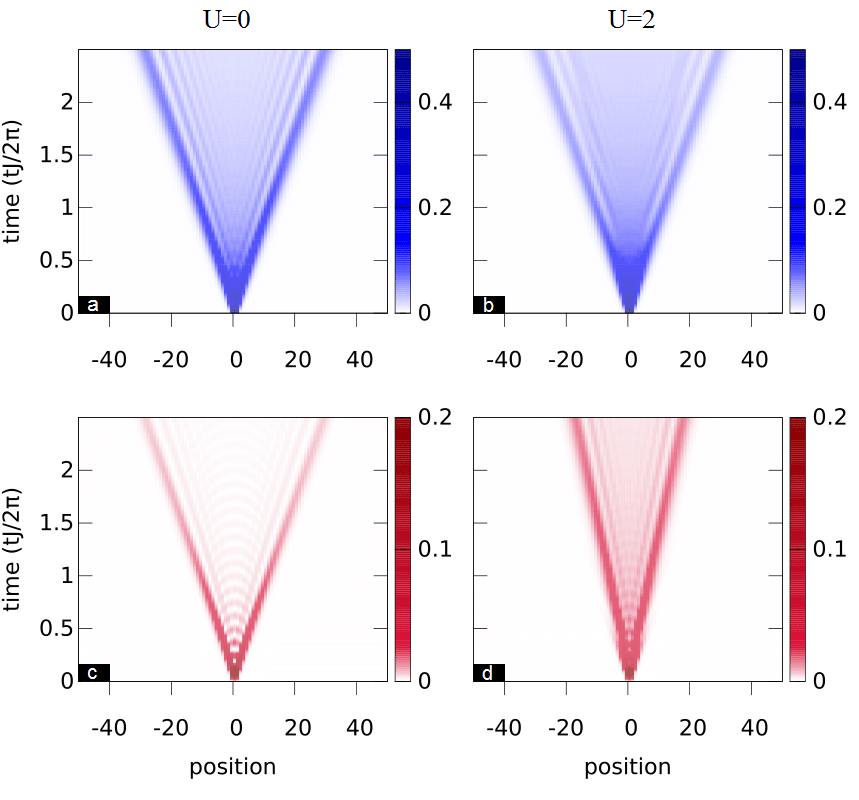}
\caption{Time evolution of different contributions $n_1(i)$ and $n_2(i)$ to the density profile $n(i)$ for different values of on-site interactions ($U=0$ and $U=2$ in left and right panel, respectively) in the case of untitled lattice. The contribution from singly occupied sites is almost insensitive to the interaction strength (upper panel with blue densities) On the other hand, time evolution of doubly occupied sites is strongly affected by interactions (bottom panel with red densities). In consequence it is responsible for appearing of the  central cone in the density distribution $n(i)$ in the presence of interactions (compare to Fig.~\ref{Fig2}).}
\label{Fig3}
\end{figure}
Fig.~\ref{Fig3} presents results for both quantities obtained for non-interacting case (left panel) and for quite strong interactions, $U=2$ (right panel). Blue and red plots refer to contributions from singly and doubly occupied sites, $n_1(i)$ and $n_2(i)$ respectively. These figures, when compared to corresponding plots in Fig. \ref{Fig2}, explicitly show that contribution from $n_2(i)$ is mostly responsible for the central cone of the complete density profile $n(i)$. It means that the quantum walk for bounded pairs are slower for stronger interactions. This result is a direct consequence of an effective tunneling rate for paired bosons \cite{Khomeriki10} $t_2=(\sqrt{U^2+16}-U)/4$ . Due to the conservation of energy, this kind of a pair-tunneling is strongly suppressed for strong interactions since intermediate state with a broken pair has essentially different energy. 

{\bf Role of long-range interactions.--}
The Bose-Hubbard Hamiltonian \eqref{hamBH} contains on-site interactions only. In consequence, the only coupling between lattice sites originates in single-particle tunnelings. Therefore, it is interesting to inspect dynamical properties of the system whenever other types of coupling are present. The simplest way to utilize this idea is to take into account long-range interactions described by Hamiltonian \eqref{Ham2b} - see also \cite{Qin2014}. 

\begin{figure}
\includegraphics[width=\columnwidth]{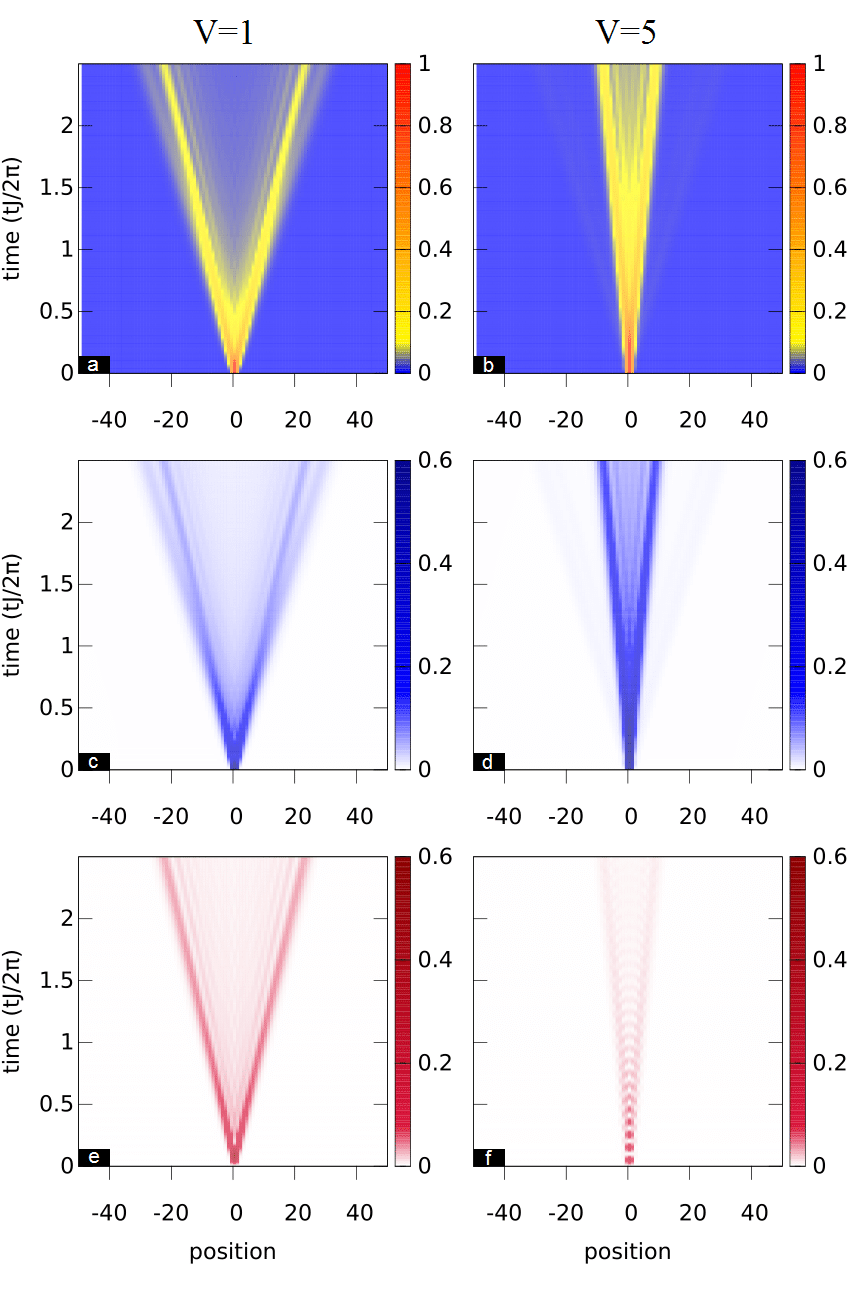}
\caption{Time evolution of the density distribution $n(i)$ and its contributions from singly and doubly occupied sites in the presence of long-range interactions. All plots are obtained for the same strength of the on-site interactions $U=2$. Left and right panels correspond to $V=1$ and $V=5$, respectively while $\alpha=3$. The comparison with the right panel in Fig.~\ref{Fig3} reveals  that for relatively weak long-range interactions a broadening  of the internal density cone (originating in doubly occupied sites) is observed while for large $V$ the transport is suppressed as the interaction modifies the effective chemical potential of different sites making the transport nonresonant. }
\label{Fig4}
\end{figure}

Fig.~\ref{Fig4} proves that the presence of long-range interactions  may significantly affect the evolution. Results from left and right panels of Fig. \ref{Fig4} should be compared with those presented in right panel of Fig.~\ref{Fig3}. Then all plots are obtained for the same initial state $|\mathtt{ini}'\rangle$ and the same on-site interaction $U=2$ but for different long-range forces (for $V=1$, and for $V=5$ with $\alpha=3$, respectively). For small $V$ its presence enhances the transport  increasing the effective tunnelings. When $V$ term becomes dominant (right panel) the fact that $V_kn_in_{i+k}$ term makes the connected sites nonresonant becomes important - the transport becomes severly slowed down, both in the single and in the double particle sector. 
 On the other hand, the exponent of the decay of interactions $\alpha$ has surprisingly negligible effect for the parameter values chosen indicating that the dominant contribution
 of this interaction comes from $Vn_in_{i+1}$ term which is $\alpha$ independent.

\begin{figure}
\includegraphics[width=\columnwidth]{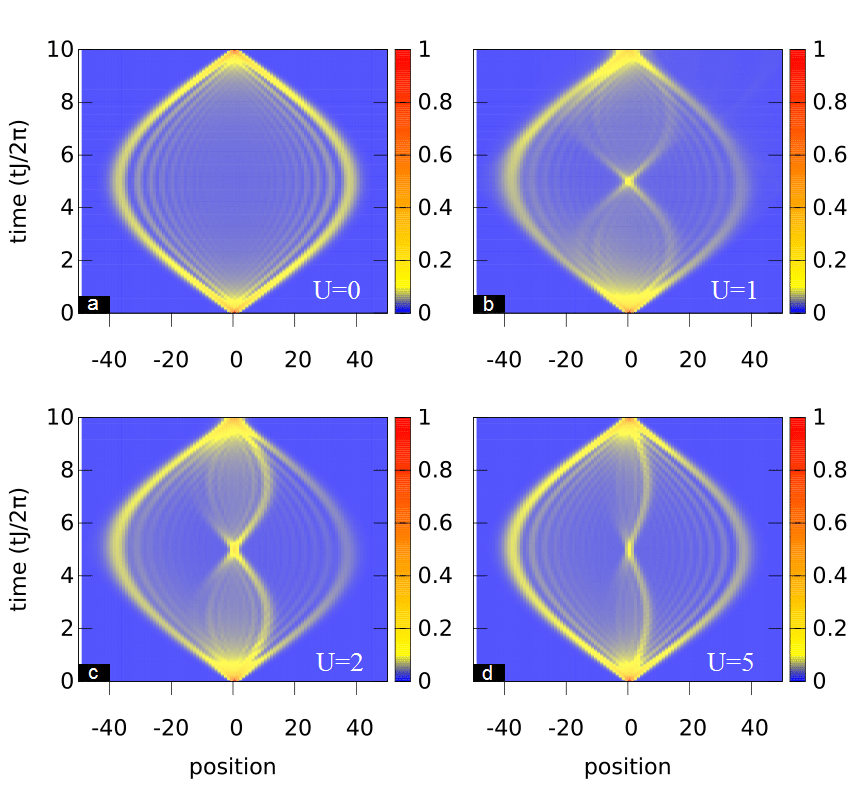}
\caption{Time evolution of the density profile $n(i)$ of interacting walkers initially occupying neighboring lattice sites in the presence of additional tilting of the lattice, $F=0.1$. For vanishing interaction the known result for the single-particle case is reproduced. Whenever local on-site interactions are present in the system, additional spatial structure in the density distribution appears. It is related to additional contribution from doubly occupied sites (compare with Fig.~\ref{Fig6}).}
\label{Fig5}
\end{figure}
\begin{figure}
\includegraphics[width=\columnwidth]{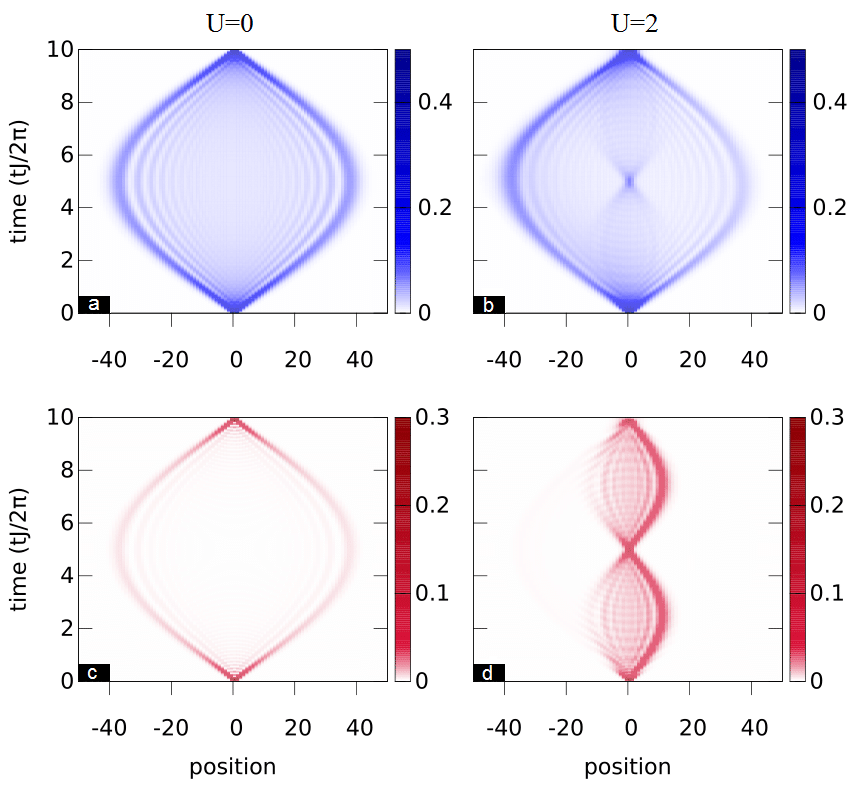}
\caption{Time evolution of different contributions to the density profile (singly and doubly occupied sites in upper and bottom panel, respectively) in the presence of external tilting of the lattice, $F=0.1$. Left and right panel are calculated for $U=0$ and $U=2$, respectively and they should be compared with appropriate plots in Fig.~\ref{Fig5}.}\label{Fig6}
\end{figure}

{\bf Effect of an external tilting.--}
Two-particle quantum walks are even more interesting when we consider non-zero external force in the Hamiltonian described by \eqref{Ham2a}. Then, as expected, Bloch oscillations are again present and  for vanishing on-site interactions previous results obtained for the single-particle case are reproduced (compare Fig.~\ref{Fig5}a and Fig.~\ref{Fig1}c). In the presence of mutual interactions an evolution of the density profile is essentially different, as presented in Fig.~\ref{Fig5}b-d. The most important change is visible in the center of the system where additional structure appears. It is a counterpart of internal cone known from $F=0$ case. As previously, non-zero on-site interactions changes mostly the density component related to doubly occupied sites. Appropriate contributions from singly and doubly occupied sites are shown in Fig.~\ref{Fig6}. It is quite interesting to note, that for stronger interactions, oscillations of the density in a doubly occupied sites have smaller width but their temporal 
period is always precisely two times shorter than the period of oscillations in the sector of singly occupied sites. These aspects  has been discussed in detail theoretically \cite{Dias07,Khomeriki10} and also confirmed in experiments \cite{Preiss2015}. 

\begin{figure}
\includegraphics[width=\columnwidth]{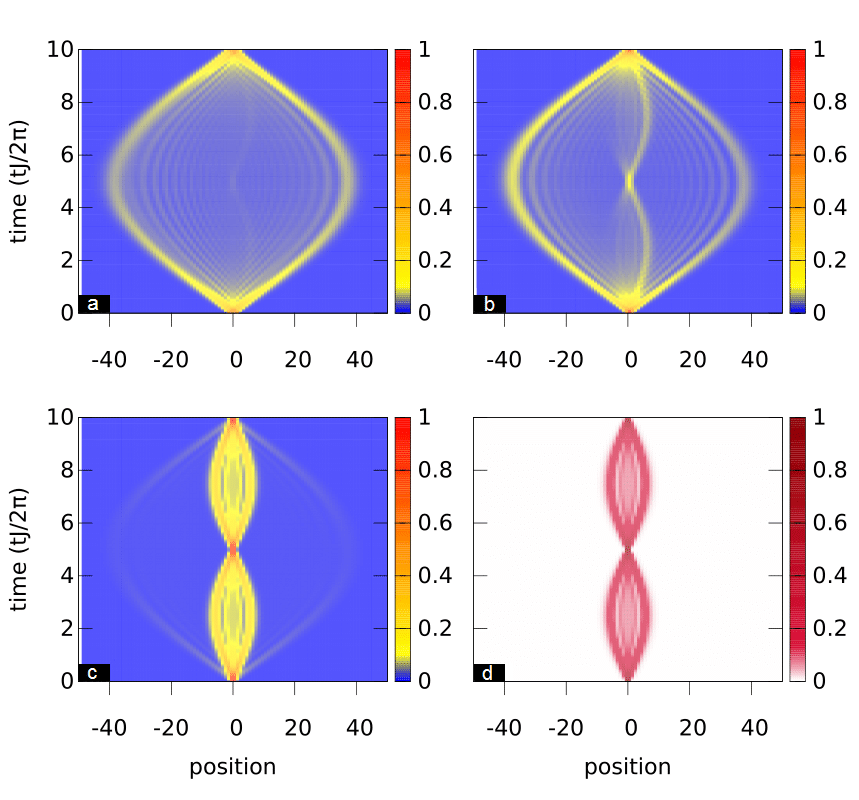}
\caption{Dependence of the time evolution of the two-particle system on the initial state. (a) Density distribution $n(i)$ when bosons initially are separated by one empty lattice site. (b) Density distribution $n(i)$ when bosons initially occupy adjacent sites. (c-d) Density distribution $n(i)$ and its contribution from doubly occupied sites $n_2(i)$ when bosons occupy exactly the same lattice site at initial moment. All plots are obtained for the same parameters, $U=5$ and $F=0.1$.}
\label{Fig7}
\end{figure}

{\bf Role of the initial state.--}
Up to now, in the case of two particles, all simulations were performed for two particles being initially in adjacent central sites. However, as known from the single-particle cases, the initial configuration strongly affects further evolution. To show that even very small change of the initial state may have huge impact to the dynamical properties of the system we compare three (almost the same from the single-particle density profile point of view)  situations. In Fig.~\ref{Fig7} we present the time evolution of the density distribution $n(i)$ in the presence of interactions and lattice tilting ($U=5$ and $F=0.1$) for three initial configurations: particles are separated by one empty site (Fig.~\ref{Fig7}a), particles occupy adjacent sites (Fig.~\ref{Fig7}b), and particles occupy exactly the same site (Fig.~\ref{Fig7}c). These examples show evidently that two-particle dynamics crucially depends on the initial state and is is affected mostly when particles start from the same or adjacent sites. For 
particles occupying initially  the same site the density distribution is carried out practically entirely in the subspace of doubly occupied sites (compare Fig.~\ref{Fig7}c and Fig.~\ref{Fig7}d). Only the minor fraction of the density distribution comes from the single-occupation sector. As it seen, the initial separation between particles leads to a rapid reduction of the importance of doubly occupied sites (Fig.~\ref{Fig7}a) and the density distribution is dominated mostly by the contribution from singly occupied sites. In consequence, the dynamics in the single-particle case is recovered.
\begin{figure}
\includegraphics[width=\columnwidth]{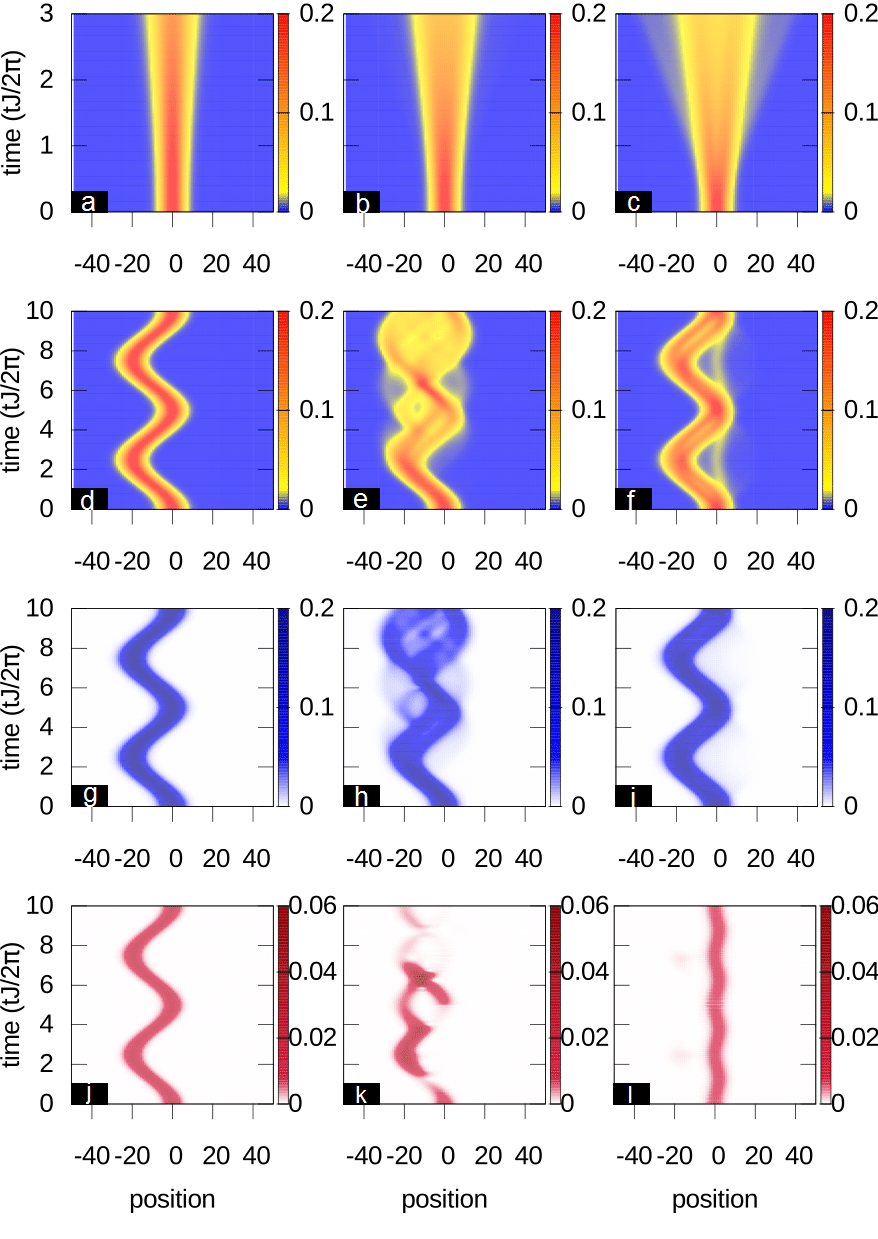} 
\caption{Time evolution of the two-particle system  for particles initially prepared in a delocalized gaussian state $|\mathtt{ini}_\sigma\rangle$ with $\sigma^2=25$. Three columns correspond to different strengths of the on-site interactions: $U =0$, $U =1$, $U =10$, respectively. First two rows show the time evolution of the density distribution $n(i)$. Top row corresponds to $F=0$ and the second row shows the results for titled lattice with $F=0.2$. Two bottom rows show contributions from singly and doubly occupied sites $n_1(i)$ and $n_2(i)$ to densities $n(i)$ presented in the second row. Although, contribution from singly occupied sites is the same for weak and strong interactions, for intermediate interactions $U\sim 1$ it is highly irregular. In this range of interactions the contribution from doubly occupied sites also undergoes a specific transition changing its oscillation frequency and its amplitude. See the main text for details. }\label{Fig8} 
\end{figure}

An observed dependence of the density distribution $n(i)$ and its contributions from doubly occupied sites $n_2(i)$ on the initial state can be viewed as a direct consequence of the conservation of the energy. Whenever on-site interactions are present, any tunneling process has to compete with the change of interaction energy between initial and final state. Whenever particles occupy the same lattice site the tunneling process breaking the pair is strongly suppressed due to the energy conservation. In consequence the second order tunneling of a whole pair become dominant and the evolution is governed mainly in the sector of doubly occupied sites $n_2(i)$. And vice versa, if particles initially occupy distant sites, interaction energy prevents system from putting both particles to the same site. Then the density distribution $n(i)$ is dominated by singly occupied sites $n_1(i)$.

Let us finally consider two-particle evolution starting from the gaussian initial distribution, $|\mathtt{ini}_\sigma\rangle$ with non-zero width, $\sigma^2=25$. Fig.~\ref{Fig8} presents the results obtained for different interactions $U$ and different tilting of the lattice $F$.   The first row (Fig.~\ref{Fig8}a-c) refers to quantum walks without lattice tilting. As it is seen, on-site interactions enhance the spreading of the initial distribution. This is not surprising bearing in mind that we consider here repulsive interactions. The second row (Fig.~\ref{Fig8}d-f) proves that interactions affect also Bloch oscillations of an initial gaussian packet when lattice tilting is present. When the strength of interactions is of the same order as tunneling rate $U=1$ (Fig.~\ref{Fig8}e) oscillations of the density profile are destroyed and some irregular behavior is observed. This is a direct consequence of a balanced competitions between single-particle tunnelings and on-site interactions. However, in strong 
interaction regime $U=10$ (Fig.~\ref{Fig8}f), regular behavior is restored and oscillations are present again. Although both evolutions of the density profile seem to be very similar for weak and strong repulsions, they are driven by fundamentally different mechanism. This difference is clearly visible when density distributions are decomposed to contributions from singly and doubly occupied sites (third and fourth row in Fig.~\ref{Fig8}, respectively). When interactions are switched off both contributions have similar evolution in time and they simply add up to the full density $n(i)$ (Fig.~\ref{Fig8}g and Fig.~\ref{Fig8}j). On the other hand, for very strong interactions, the contribution from singly occupied sites remain unchanged but contribution from doublons is essentially different, i.e., amplitude of its oscillations is much smaller and frequency is doubled (Fig.~\ref{Fig8}i and Fig.~\ref{Fig8}l). At the same time the doublon contribution to the dominant part of the density profile (Fig.~\ref{Fig8}f) 
is negligible. This observation suggests that the system undergoes specific transition between these two scenarios, i.e., the same evolution of both density contributions for weak interactions and completely different evolution of these contributions for strong interactions. For intermediate interactions system is strongly affected by both scenarios which is manifested in irregular evolution of the densities (Fig.~\ref{Fig8}e, Fig.~\ref{Fig8}h, and Fig.~\ref{Fig8}k). This observation can be utilised experimentally when different two-body correlations are studied \cite{Preiss2015}.

\begin{figure}
\includegraphics[width=\columnwidth]{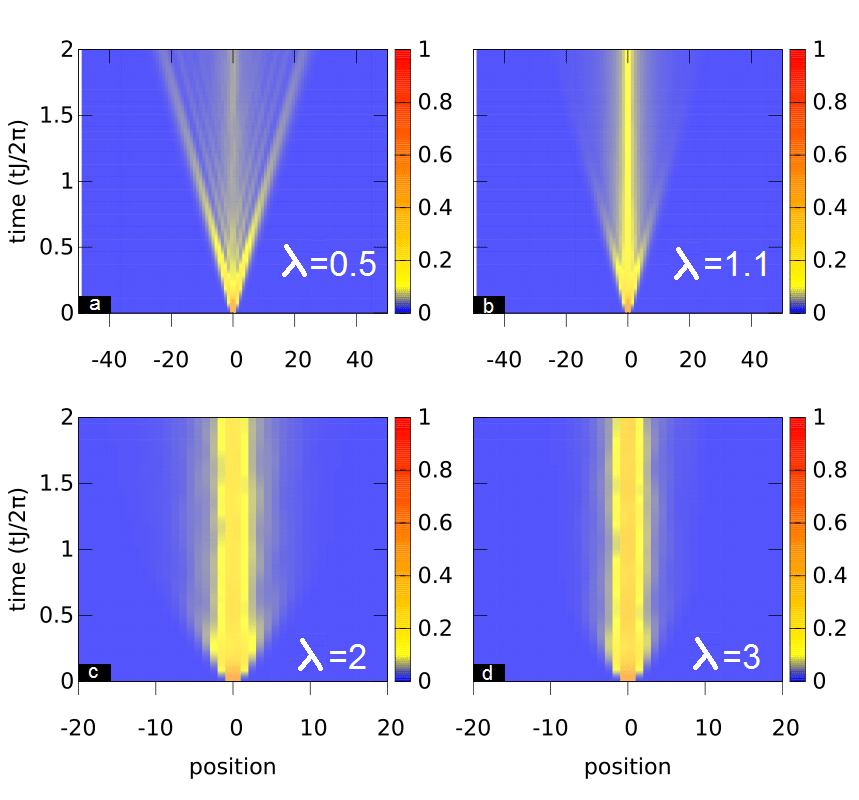}
\caption{Time evolution of the density distribution $n(i)$ for a single quantum walker initially occupying single lattice site in the presence of the lattice disorder. For large enough disorder ($\lambda>2$) system manifests the Aubry-Andr\'e localization.}
\label{Fig9}
\end{figure} 
\section{Disordered systems}\label{AN}
Let us now study what information we may obtain from the quantum walkers approach in the case of disordered systems. To that end we slightly modify the Hamiltonian and we add a small on-site disorder term \eqref{Ham2c} of the form
\begin{equation}
\hat{D} = \lambda \sum_{i=1} \cos\left[2 \pi (\tau i +\phi)\right] \hat a_i^{\dagger}  \hat a_i,
\end{equation} 
where $\lambda$ measures a strength of a disorder. Instead of a truly random disorder, we investigate the quasi-random disorder induced by a cosine modulation of on-site energies (chemical potential). Such a situation is routinely realized in experiments \cite{Schreiber2015,Bordia2016,Bordia2017} by adding a second weak optical lattice with a period almost incommensurate with the primary lattice. Here, we fix $\tau=(\sqrt{5}-1)/2$ while $\phi$ is an arbitrary but fixed for a given realization. The results obtained are averaged over various realizations (typically few thousands) obtained by varying the phase $\phi$.   

{\bf Single-particle localization.--}
Whenever one considers evolution of a single quantum particle in not tilted lattice ($F=0$) the situation is well known and understood. Therefore, we will only briefly show its properties. 

Without going into details, it is a matter of fact that any one-dimensional system of this type manifests Aubry-Andr\'e \cite{AA80} localization in the configuration space for $\lambda>2$ \cite{Mueller2009} rather than technically different Anderson localization. The latter occurs for a truly random disorder for all eigenstates in one-dimensional systems regardless the disorder strength.
In Fig.~\ref{Fig9} the time evolution of the density distribution for different disorder values are presented. A weak disorder $\lambda=0.5$ (Fig.~\ref{Fig9}a) leaves the early time evolution almost unaltered. When compare with $\lambda=0$ case (Fig.~\ref{Fig1}a) one observes almost purely ballistic expansion with some small corrections. Note however, that for stronger disorders the transport across the lattice is slown down and a central part of the wave packet seems to be trapped close to the site initially occupied (Fig.~\ref{Fig9}b). Situation is markedly different for $\lambda=2$ (the critical value) as well as $\lambda=3$ when after a short initial spread the wave packet freezes its position in time. This phenomenon is one of the signatures of the localization.

{\bf Many-body localization in an absence of tilting.--}
\begin{figure}
\includegraphics[width=\columnwidth]{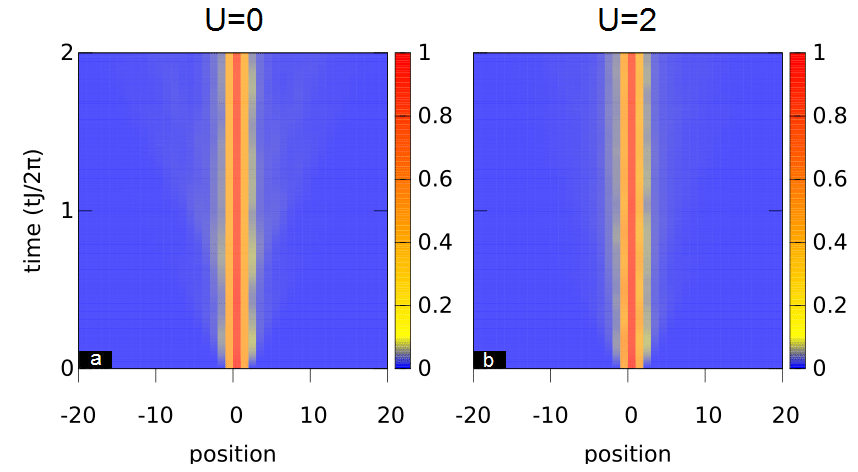}
\caption{The time evolution of the density distribution $n(i)$ for two bosons initially occupying adjacent sites of the lattice in the presence of disorder ($\lambda=2$). As it is seen, in the presence of interactions (right panel) the propagating wave packet is localized better than in the absence of interactions (left panel). Note, that the localization of the wave packet is amplified although interactions are repulsive.} 
 \label{Fig10} 
\end{figure}
While quantum walk of two noninteracting particles in disordered potential was considered in the framework of Anderson localization \cite{Lahini2010}
in the case of interacting particles situation is much more complicated. Our understanding of physics in this case in the presence of the disorder underwent significant progress recently mostly due to identification of the many-body localization phenomenon \cite{Basko06}. Previously, common understanding was build on the assumption that interacting particles in the presence of the disorder should ``thermalize'' in the sense of eigenvector thermalisation hypothesis 
\cite{srednicki94}. While the isolated system as a whole evolves in a unitary way without loosing any information, averages of local observables in typical evolved state should thermalize, i.e., the system looses locally whole memory about the initial state. The many-body localization is completely opposite effect. In the presence of localization local averages do not thermalize, systems
are conjectured to be integrable and possessing a complete set of local integrals of motion \cite{Serbyn2013}. Many-body localization has been extensively studied and in in last five years many interesting results were obtained. For excellent recent reviews see e.g. \cite{Huse14,Rahul15}.
\begin{figure}
\includegraphics[width=\columnwidth]{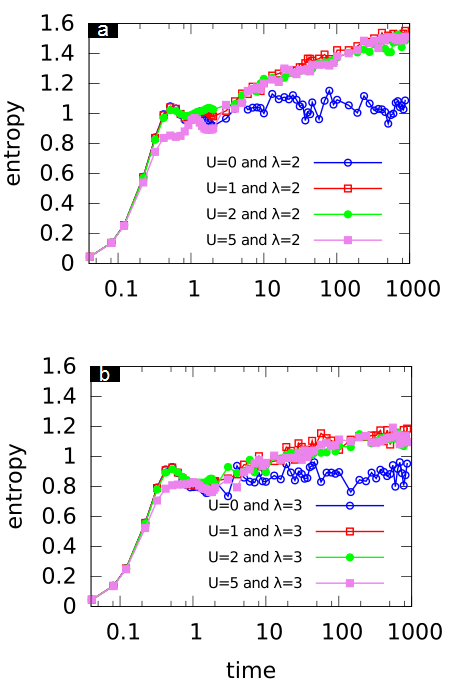}
\caption{Time dependence of the entanglement entropy ${\cal S}_A$ for different strengths of the on-site interactions and for two different disorder amplitudes $\lambda=2$ (upper panel) and $\lambda=3$ (bottom panel). One observes that, in both cases of the disorder the entanglement entropy saturates for noninteracting system $U=0$. Whenever interactions between particles are present the entanglement entropy grows logarithmically in time with the slope almost independent on the interaction strength. This observation is in agreement with predictions for the many-body localization phenomena.}
 \label{Fig11} 
\end{figure}

On a theoretical level, many-body localization is very often build on identifying different properties of the system in the thermodynamic limit. Recently, a few existing experiments consider finite but quite large systems to support this approach \cite{Schreiber2015,Bordia2016,Bordia2017}. Here, we address and we try to investigate possible answers to other interesting question: which properties of a many-body system may be correctly captured and identified for just two particles using quantum walk approach.

To place an answer in a proper context let us compare localization properties of the system of two bosons with those obtained for single particle subjected to the same disorder of the lattice. The time-evolution of the system for exemplary parameters is shown in Fig.~\ref{Fig10}. Again we consider the situation where at the initial moment bosons occupy adjacent sites of the lattice. During the evolution particles interfere with the tendency to localize (the disorder amplitude $\lambda=2$ is critical for non interacting particles). Observe that in the presence of interactions (right panel) the propagating wave packet seems to localize better than in the absence of interactions (left panel).

This behavior can be quantified in the spirit of the many-body localization phenomenon. One of the key characteristic features of the many-body localization making it fundamentally different from any single-particle model is that the entanglement entropy between two subsystems of the model grows logarithmically in time \cite{Znidaric2008,Bardarson2012}.
In order to compute the entanglement entropy we divide our lattice into two equal sublattices $A$ and $B$ and we compute the reduced density matrix of the subsystem by tracing-out remaining degrees of freedom from the density matrix of the system
\begin{equation} 
\rho_A(t)=\mathrm{Tr}_B\left(|\psi(t)\rangle\langle\psi(t)|\right).
\end{equation}
Then we directly determine the entanglement entropy as
\begin{equation}
{\cal S}_A(t)=-\mathrm{Tr}\left[\,\rho_A(t)\,\log\rho_A(t)\right].
\end{equation}
Indeed, as seen Fig.~\ref{Fig11}, a logarithmic growth of the entanglement entropy is observed as soon as the on-site interactions have non-zero values. It seems interesting that this fundamental signature of the many-body localization, associated typically with many-body physics, appears here as a characteristics of two-particle dynamics. This raises the question if the logarithmic entropy growth observed is really a signature of the many-body localization or rather it is a feature of two particle entanglement in the presence of disorder. The answer is straightforward when non-interacting cases of two bosons are considered. As it is seen in Fig.~\ref{Fig11} (blue lines) in these cases we observe saturation of the entanglement entropy, which is in agreement with its standard behavior for Anderson localized phase rather than for the many-body localization phenomenon.

{\bf Localization in a tilted lattice.--}
\begin{figure}
\includegraphics[width=\columnwidth]{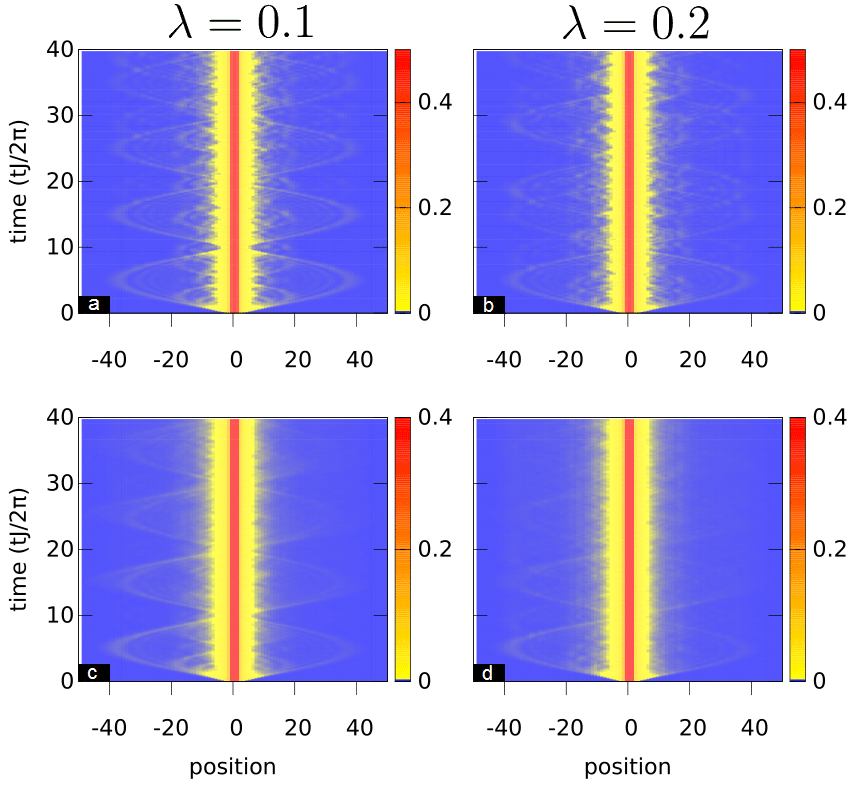}
\caption{Time evolution of the density distribution $n(i)$ of two bosonic walkers initially occupying adjacent sites in the presence of the disorder when additional tilting of the lattice is turned on. Top and bottom rows correspond to non-interacting case $U=0$ and on-site interaction $U=2$, respectively. Results are averaged over 100 realizations of disorder. Note, that in the presence of interaction specific decay of the Bloch oscillations is present.}
\label{Fig12}
\end{figure}
\begin{figure}
\includegraphics[width=\columnwidth]{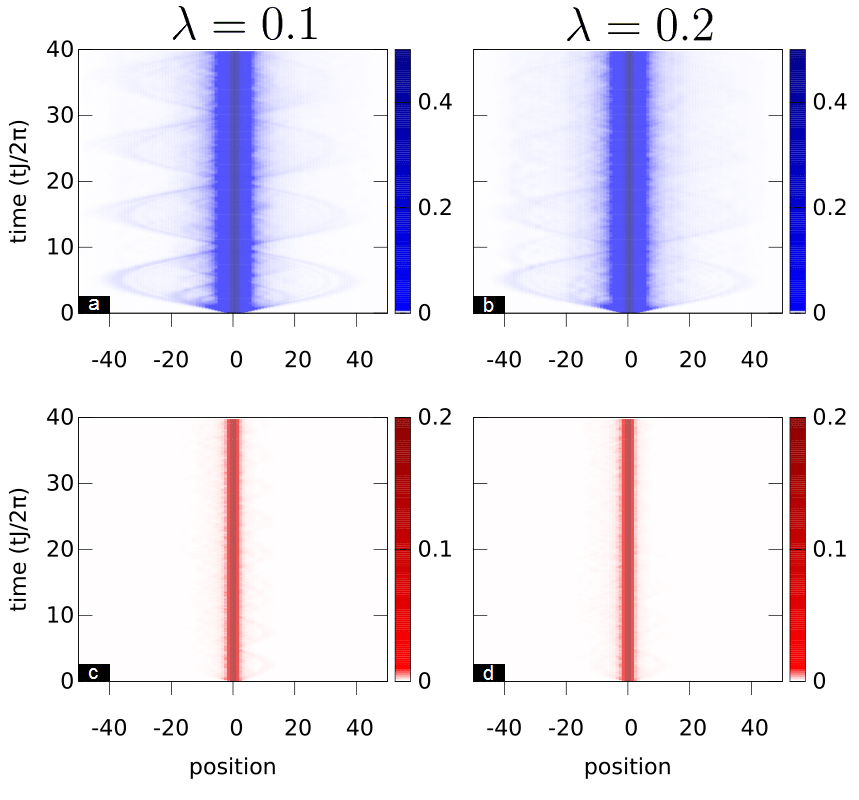}
\caption{Time evolution of contributions from singly and doubly occupied sites $n_1(i)$ and $n_2(i)$ to the density distribution in the case of interacting system $U=2$ presented in Fig.~\ref{Fig13}.}
\label{Fig13}
\end{figure}
Finally, let us present results for the evolution in tilted lattices in the presence of the quasi-random disorder. We consider small disorder amplitudes, far from the localization, i.e., $\lambda=0.1$ and $\lambda=0.2$. We checked that the results obtained for these parameters are generic and similar for other values of $\lambda$. 

As before we start with two particles occupying adjacent sites in the centre of the lattice. In the presence of the disorder Bloch oscillations are still present but they have irregular character and additional decay of their amplitude is observed (Fig.~\ref{Fig12}). The decay is larger for higher disorder present in the lattice. It is worth noticing, that similar damping of Bloch oscillations was observed recently in experiment with oscillating electrons \cite{Lyssenko97}. It has been also studied theoretically \cite{Diez96,Schulte2008}. Note, that the damping of Bloch oscillations is enhanced by interactions.

The collective properties of the system in the presence of an external field $F$ and interactions $U$ may be further analyzed by decomposition of the density profile $n(i)$ to its occupation component $n_1(i)$ and $n_2(i)$ (Fig.~\ref{Fig13}). As it is clearly visible, in contrast to the case without disorder, the density of doubly occupied sites is almost completely pined down to the area around initially populated sites and oscillations of the density in this sector are almost invisible. Surprisingly, this strong dephasing effect is present for relatively weak disorder $\lambda=0.1$. This suggest, that a role of disorder can be substantially amplified when many-body problems are considered. 

\section{Conclusions}\label{Concl}
We have shown numerical results for two bosonic quantum walkers in optical lattices with external field in a wide range of parameters. The additional field term leads to Bloch oscillations which temporal period and width strongly depend on a field strength. In a two-particle system on-site interactions induce oscillations of the density of doubly occupied sites at double frequency with appropriately diminished amplitude. In the case of initially delocalized particles (with gaussian distribution) the interactions tend to destabilize oscillations for intermediate interaction values. Surprisingly, stronger interactions again stabilize Bloch oscillations of the wave packet. 

In addition, we have also analyzed behavior of the quantum walkers in the presence of lattice disorder. We believe that such studies have been missing in the literature. Interestingly we have shown that the logarithmic growth in time of the entanglement entropy, which is characteristic for the many-body localization, may be observed already on the level of  quantum walks of two particles. Moreover, even a relatively small tilting of the lattice strongly diminishes a role of the disorder. The system remains localized with Bloch oscillations being damped.

Since, different scenarios of quantum walks are accessible in nowadays experiments on ultra-cold atoms \cite{Preiss2015}, we believe that the results presented may have some importance for further studies of these kind of systems.

\begin{acknowledgments} 
The authors are grateful to Mariusz Gajda for his lasting willingness for discussions on various topics not always related to physics. This work was performed within the  EU Horizon2020 FET project QUIC (no. 641122). We acknowledge support of the (Polish) National Science Centre via projects No 2015/19/B/ST2/01028 (DW), 2016/22/E/ST2/00555 (TS), and 2016/21/B/ST2/01086 (JZ). 
\end{acknowledgments}


\begin{thebibliography}{57}%
\makeatletter
\providecommand \@ifxundefined [1]{%
 \@ifx{#1\undefined}
}%
\providecommand \@ifnum [1]{%
 \ifnum #1\expandafter \@firstoftwo
 \else \expandafter \@secondoftwo
 \fi
}%
\providecommand \@ifx [1]{%
 \ifx #1\expandafter \@firstoftwo
 \else \expandafter \@secondoftwo
 \fi
}%
\providecommand \natexlab [1]{#1}%
\providecommand \enquote  [1]{``#1''}%
\providecommand \bibnamefont  [1]{#1}%
\providecommand \bibfnamefont [1]{#1}%
\providecommand \citenamefont [1]{#1}%
\providecommand \href@noop [0]{\@secondoftwo}%
\providecommand \href [0]{\begingroup \@sanitize@url \@href}%
\providecommand \@href[1]{\@@startlink{#1}\@@href}%
\providecommand \@@href[1]{\endgroup#1\@@endlink}%
\providecommand \@sanitize@url [0]{\catcode `\\12\catcode `\$12\catcode
  `\&12\catcode `\#12\catcode `\^12\catcode `\_12\catcode `\%12\relax}%
\providecommand \@@startlink[1]{}%
\providecommand \@@endlink[0]{}%
\providecommand \url  [0]{\begingroup\@sanitize@url \@url }%
\providecommand \@url [1]{\endgroup\@href {#1}{\urlprefix }}%
\providecommand \urlprefix  [0]{URL }%
\providecommand \Eprint [0]{\href }%
\providecommand \doibase [0]{http://dx.doi.org/}%
\providecommand \selectlanguage [0]{\@gobble}%
\providecommand \bibinfo  [0]{\@secondoftwo}%
\providecommand \bibfield  [0]{\@secondoftwo}%
\providecommand \translation [1]{[#1]}%
\providecommand \BibitemOpen [0]{}%
\providecommand \bibitemStop [0]{}%
\providecommand \bibitemNoStop [0]{.\EOS\space}%
\providecommand \EOS [0]{\spacefactor3000\relax}%
\providecommand \BibitemShut  [1]{\csname bibitem#1\endcsname}%
\let\auto@bib@innerbib\@empty
\bibitem [{\citenamefont {Manouchehri}\ and\ \citenamefont
  {Wang}(2014)}]{QuantumWalks}%
  \BibitemOpen
  \bibfield  {author} {\bibinfo {author} {\bibfnamefont {K.}~\bibnamefont
  {Manouchehri}}\ and\ \bibinfo {author} {\bibfnamefont {J.}~\bibnamefont
  {Wang}},\ }\href@noop {} {\emph {\bibinfo {title} {{Physical Implementation
  of Quantum Walks}}}}\ (\bibinfo  {publisher} {{Springer-Verlag}},\ \bibinfo
  {address} {{Berlin, Heidelberg}},\ \bibinfo {year} {2014})\BibitemShut
  {NoStop}%
\bibitem [{\citenamefont {Mendez}\ and\ \citenamefont
  {Bastard}(1993)}]{Mendez93}%
  \BibitemOpen
  \bibfield  {author} {\bibinfo {author} {\bibfnamefont {E.}~\bibnamefont
  {Mendez}}\ and\ \bibinfo {author} {\bibfnamefont {G.}~\bibnamefont
  {Bastard}},\ }\href {\doibase 10.1103/1.881353} {\bibfield  {journal}
  {\bibinfo  {journal} {Phys. Today}\ }\textbf {\bibinfo {volume} {46}},\
  \bibinfo {pages} {34} (\bibinfo {year} {1993})}\BibitemShut {NoStop}%
\bibitem [{\citenamefont {Lewenstein}\ \emph {et~al.}(2007)\citenamefont
  {Lewenstein}, \citenamefont {Sanpera}, \citenamefont {Ahufinger},
  \citenamefont {Damski}, \citenamefont {Sen(De)},\ and\ \citenamefont
  {Sen}}]{Lewenstein07}%
  \BibitemOpen
  \bibfield  {author} {\bibinfo {author} {\bibfnamefont {M.}~\bibnamefont
  {Lewenstein}}, \bibinfo {author} {\bibfnamefont {A.}~\bibnamefont {Sanpera}},
  \bibinfo {author} {\bibfnamefont {V.}~\bibnamefont {Ahufinger}}, \bibinfo
  {author} {\bibfnamefont {B.}~\bibnamefont {Damski}}, \bibinfo {author}
  {\bibfnamefont {A.}~\bibnamefont {Sen(De)}}, \ and\ \bibinfo {author}
  {\bibfnamefont {U.}~\bibnamefont {Sen}},\ }\href@noop {} {\bibfield
  {journal} {\bibinfo  {journal} {Adv. Phys.}\ }\textbf {\bibinfo {volume}
  {56}},\ \bibinfo {pages} {243} (\bibinfo {year} {2007})}\BibitemShut
  {NoStop}%
\bibitem [{\citenamefont {Lewenstein}\ \emph {et~al.}(2012)\citenamefont
  {Lewenstein}, \citenamefont {Sanpera},\ and\ \citenamefont
  {Ahufinger}}]{Lewenstein12}%
  \BibitemOpen
  \bibfield  {author} {\bibinfo {author} {\bibfnamefont {M.}~\bibnamefont
  {Lewenstein}}, \bibinfo {author} {\bibfnamefont {A.}~\bibnamefont {Sanpera}},
  \ and\ \bibinfo {author} {\bibfnamefont {V.}~\bibnamefont {Ahufinger}},\
  }\href@noop {} {\emph {\bibinfo {title} {Ultracold Atoms in Optical Lattices:
  Simulating Many-Body Quantum Systems}}}\ (\bibinfo  {publisher} {Oxford
  University Press},\ \bibinfo {year} {2012})\BibitemShut {NoStop}%
\bibitem [{\citenamefont {Ronzheimer}\ \emph {et~al.}(2013)\citenamefont
  {Ronzheimer}, \citenamefont {Schreiber}, \citenamefont {Braun}, \citenamefont
  {Hodgman}, \citenamefont {Langer}, \citenamefont {McCulloch}, \citenamefont
  {Heidrich-Meisner}, \citenamefont {Bloch},\ and\ \citenamefont
  {Schneider}}]{Ronzheimer2013}%
  \BibitemOpen
  \bibfield  {author} {\bibinfo {author} {\bibfnamefont {J.~P.}\ \bibnamefont
  {Ronzheimer}}, \bibinfo {author} {\bibfnamefont {M.}~\bibnamefont
  {Schreiber}}, \bibinfo {author} {\bibfnamefont {S.}~\bibnamefont {Braun}},
  \bibinfo {author} {\bibfnamefont {S.~S.}\ \bibnamefont {Hodgman}}, \bibinfo
  {author} {\bibfnamefont {S.}~\bibnamefont {Langer}}, \bibinfo {author}
  {\bibfnamefont {I.~P.}\ \bibnamefont {McCulloch}}, \bibinfo {author}
  {\bibfnamefont {F.}~\bibnamefont {Heidrich-Meisner}}, \bibinfo {author}
  {\bibfnamefont {I.}~\bibnamefont {Bloch}}, \ and\ \bibinfo {author}
  {\bibfnamefont {U.}~\bibnamefont {Schneider}},\ }\href {\doibase
  10.1103/PhysRevLett.110.205301} {\bibfield  {journal} {\bibinfo  {journal}
  {Phys. Rev. Lett.}\ }\textbf {\bibinfo {volume} {110}},\ \bibinfo {pages}
  {205301} (\bibinfo {year} {2013})}\BibitemShut {NoStop}%
\bibitem [{\citenamefont {Sorg}\ \emph {et~al.}(2014)\citenamefont {Sorg},
  \citenamefont {Vidmar}, \citenamefont {Pollet},\ and\ \citenamefont
  {Heidrich-Meisner}}]{Song2014}%
  \BibitemOpen
  \bibfield  {author} {\bibinfo {author} {\bibfnamefont {S.}~\bibnamefont
  {Sorg}}, \bibinfo {author} {\bibfnamefont {L.}~\bibnamefont {Vidmar}},
  \bibinfo {author} {\bibfnamefont {L.}~\bibnamefont {Pollet}}, \ and\ \bibinfo
  {author} {\bibfnamefont {F.}~\bibnamefont {Heidrich-Meisner}},\ }\href
  {\doibase 10.1103/PhysRevA.90.033606} {\bibfield  {journal} {\bibinfo
  {journal} {Phys. Rev. A}\ }\textbf {\bibinfo {volume} {90}},\ \bibinfo
  {pages} {033606} (\bibinfo {year} {2014})}\BibitemShut {NoStop}%
\bibitem [{\citenamefont {Vidmar}\ \emph {et~al.}(2015)\citenamefont {Vidmar},
  \citenamefont {Ronzheimer}, \citenamefont {Schreiber}, \citenamefont {Braun},
  \citenamefont {Hodgman}, \citenamefont {Langer}, \citenamefont
  {Heidrich-Meisner}, \citenamefont {Bloch},\ and\ \citenamefont
  {Schneider}}]{Vidmar2015}%
  \BibitemOpen
  \bibfield  {author} {\bibinfo {author} {\bibfnamefont {L.}~\bibnamefont
  {Vidmar}}, \bibinfo {author} {\bibfnamefont {J.~P.}\ \bibnamefont
  {Ronzheimer}}, \bibinfo {author} {\bibfnamefont {M.}~\bibnamefont
  {Schreiber}}, \bibinfo {author} {\bibfnamefont {S.}~\bibnamefont {Braun}},
  \bibinfo {author} {\bibfnamefont {S.~S.}\ \bibnamefont {Hodgman}}, \bibinfo
  {author} {\bibfnamefont {S.}~\bibnamefont {Langer}}, \bibinfo {author}
  {\bibfnamefont {F.}~\bibnamefont {Heidrich-Meisner}}, \bibinfo {author}
  {\bibfnamefont {I.}~\bibnamefont {Bloch}}, \ and\ \bibinfo {author}
  {\bibfnamefont {U.}~\bibnamefont {Schneider}},\ }\href {\doibase
  10.1103/PhysRevLett.115.175301} {\bibfield  {journal} {\bibinfo  {journal}
  {Phys. Rev. Lett.}\ }\textbf {\bibinfo {volume} {115}},\ \bibinfo {pages}
  {175301} (\bibinfo {year} {2015})}\BibitemShut {NoStop}%
\bibitem [{\citenamefont {Hauschild}\ \emph {et~al.}(2015)\citenamefont
  {Hauschild}, \citenamefont {Pollmann},\ and\ \citenamefont
  {Heidrich-Meisner}}]{Hauschild2015}%
  \BibitemOpen
  \bibfield  {author} {\bibinfo {author} {\bibfnamefont {J.}~\bibnamefont
  {Hauschild}}, \bibinfo {author} {\bibfnamefont {F.}~\bibnamefont {Pollmann}},
  \ and\ \bibinfo {author} {\bibfnamefont {F.}~\bibnamefont
  {Heidrich-Meisner}},\ }\href {\doibase 10.1103/PhysRevA.92.053629} {\bibfield
   {journal} {\bibinfo  {journal} {Phys. Rev. A}\ }\textbf {\bibinfo {volume}
  {92}},\ \bibinfo {pages} {053629} (\bibinfo {year} {2015})}\BibitemShut
  {NoStop}%
\bibitem [{\citenamefont {Peruzzo}\ \emph {et~al.}(2010)\citenamefont
  {Peruzzo}, \citenamefont {Lobino}, \citenamefont {Matthews}, \citenamefont
  {Matsuda}, \citenamefont {Politi}, \citenamefont {Poulios}, \citenamefont
  {Zhou}, \citenamefont {Lahini}, \citenamefont {Ismail}, \citenamefont
  {W{\"o}rhoff}, \citenamefont {Bromberg}, \citenamefont {Silberberg},
  \citenamefont {Thompson},\ and\ \citenamefont {OBrien}}]{Peruzzo2010}%
  \BibitemOpen
  \bibfield  {author} {\bibinfo {author} {\bibfnamefont {A.}~\bibnamefont
  {Peruzzo}}, \bibinfo {author} {\bibfnamefont {M.}~\bibnamefont {Lobino}},
  \bibinfo {author} {\bibfnamefont {J.~C.~F.}\ \bibnamefont {Matthews}},
  \bibinfo {author} {\bibfnamefont {N.}~\bibnamefont {Matsuda}}, \bibinfo
  {author} {\bibfnamefont {A.}~\bibnamefont {Politi}}, \bibinfo {author}
  {\bibfnamefont {K.}~\bibnamefont {Poulios}}, \bibinfo {author} {\bibfnamefont
  {X.-Q.}\ \bibnamefont {Zhou}}, \bibinfo {author} {\bibfnamefont
  {Y.}~\bibnamefont {Lahini}}, \bibinfo {author} {\bibfnamefont
  {N.}~\bibnamefont {Ismail}}, \bibinfo {author} {\bibfnamefont
  {K.}~\bibnamefont {W{\"o}rhoff}}, \bibinfo {author} {\bibfnamefont
  {Y.}~\bibnamefont {Bromberg}}, \bibinfo {author} {\bibfnamefont
  {Y.}~\bibnamefont {Silberberg}}, \bibinfo {author} {\bibfnamefont {M.~G.}\
  \bibnamefont {Thompson}}, \ and\ \bibinfo {author} {\bibfnamefont {J.~L.}\
  \bibnamefont {OBrien}},\ }\href {\doibase 10.1126/science.1193515} {\bibfield
   {journal} {\bibinfo  {journal} {Science}\ }\textbf {\bibinfo {volume}
  {329}},\ \bibinfo {pages} {1500} (\bibinfo {year} {2010})},\ \Eprint
  {http://arxiv.org/abs/http://science.sciencemag.org/content/329/5998/1500.full.pdf}
  {http://science.sciencemag.org/content/329/5998/1500.full.pdf} \BibitemShut
  {NoStop}%
\bibitem [{\citenamefont {Bromberg}\ \emph {et~al.}(2009)\citenamefont
  {Bromberg}, \citenamefont {Lahini}, \citenamefont {Morandotti},\ and\
  \citenamefont {Silberberg}}]{Bromberg2009}%
  \BibitemOpen
  \bibfield  {author} {\bibinfo {author} {\bibfnamefont {Y.}~\bibnamefont
  {Bromberg}}, \bibinfo {author} {\bibfnamefont {Y.}~\bibnamefont {Lahini}},
  \bibinfo {author} {\bibfnamefont {R.}~\bibnamefont {Morandotti}}, \ and\
  \bibinfo {author} {\bibfnamefont {Y.}~\bibnamefont {Silberberg}},\ }\href
  {\doibase 10.1103/PhysRevLett.102.253904} {\bibfield  {journal} {\bibinfo
  {journal} {Phys. Rev. Lett.}\ }\textbf {\bibinfo {volume} {102}},\ \bibinfo
  {pages} {253904} (\bibinfo {year} {2009})}\BibitemShut {NoStop}%
\bibitem [{\citenamefont {Lahini}\ \emph {et~al.}(2010)\citenamefont {Lahini},
  \citenamefont {Bromberg}, \citenamefont {Christodoulides},\ and\
  \citenamefont {Silberberg}}]{Lahini2010}%
  \BibitemOpen
  \bibfield  {author} {\bibinfo {author} {\bibfnamefont {Y.}~\bibnamefont
  {Lahini}}, \bibinfo {author} {\bibfnamefont {Y.}~\bibnamefont {Bromberg}},
  \bibinfo {author} {\bibfnamefont {D.~N.}\ \bibnamefont {Christodoulides}}, \
  and\ \bibinfo {author} {\bibfnamefont {Y.}~\bibnamefont {Silberberg}},\
  }\href {\doibase 10.1103/PhysRevLett.105.163905} {\bibfield  {journal}
  {\bibinfo  {journal} {Phys. Rev. Lett.}\ }\textbf {\bibinfo {volume} {105}},\
  \bibinfo {pages} {163905} (\bibinfo {year} {2010})}\BibitemShut {NoStop}%
\bibitem [{\citenamefont {Lahini}\ \emph {et~al.}(2012)\citenamefont {Lahini},
  \citenamefont {Verbin}, \citenamefont {Huber}, \citenamefont {Bromberg},
  \citenamefont {Pugatch},\ and\ \citenamefont {Silberberg}}]{Lahini2012}%
  \BibitemOpen
  \bibfield  {author} {\bibinfo {author} {\bibfnamefont {Y.}~\bibnamefont
  {Lahini}}, \bibinfo {author} {\bibfnamefont {M.}~\bibnamefont {Verbin}},
  \bibinfo {author} {\bibfnamefont {S.~D.}\ \bibnamefont {Huber}}, \bibinfo
  {author} {\bibfnamefont {Y.}~\bibnamefont {Bromberg}}, \bibinfo {author}
  {\bibfnamefont {R.}~\bibnamefont {Pugatch}}, \ and\ \bibinfo {author}
  {\bibfnamefont {Y.}~\bibnamefont {Silberberg}},\ }\href {\doibase
  10.1103/PhysRevA.86.011603} {\bibfield  {journal} {\bibinfo  {journal} {Phys.
  Rev. A}\ }\textbf {\bibinfo {volume} {86}},\ \bibinfo {pages} {011603}
  (\bibinfo {year} {2012})}\BibitemShut {NoStop}%
\bibitem [{\citenamefont {Solntsev}\ \emph {et~al.}(2012)\citenamefont
  {Solntsev}, \citenamefont {Sukhorukov}, \citenamefont {Neshev},\ and\
  \citenamefont {Kivshar}}]{Solntsev2012}%
  \BibitemOpen
  \bibfield  {author} {\bibinfo {author} {\bibfnamefont {A.~S.}\ \bibnamefont
  {Solntsev}}, \bibinfo {author} {\bibfnamefont {A.~A.}\ \bibnamefont
  {Sukhorukov}}, \bibinfo {author} {\bibfnamefont {D.~N.}\ \bibnamefont
  {Neshev}}, \ and\ \bibinfo {author} {\bibfnamefont {Y.~S.}\ \bibnamefont
  {Kivshar}},\ }\href {\doibase 10.1103/PhysRevLett.108.023601} {\bibfield
  {journal} {\bibinfo  {journal} {Phys. Rev. Lett.}\ }\textbf {\bibinfo
  {volume} {108}},\ \bibinfo {pages} {023601} (\bibinfo {year}
  {2012})}\BibitemShut {NoStop}%
\bibitem [{\citenamefont {Benedetti}\ \emph {et~al.}(2012)\citenamefont
  {Benedetti}, \citenamefont {Buscemi},\ and\ \citenamefont
  {Bordone}}]{Benedetti2012}%
  \BibitemOpen
  \bibfield  {author} {\bibinfo {author} {\bibfnamefont {C.}~\bibnamefont
  {Benedetti}}, \bibinfo {author} {\bibfnamefont {F.}~\bibnamefont {Buscemi}},
  \ and\ \bibinfo {author} {\bibfnamefont {P.}~\bibnamefont {Bordone}},\ }\href
  {\doibase 10.1103/PhysRevA.85.042314} {\bibfield  {journal} {\bibinfo
  {journal} {Phys. Rev. A}\ }\textbf {\bibinfo {volume} {85}},\ \bibinfo
  {pages} {042314} (\bibinfo {year} {2012})}\BibitemShut {NoStop}%
\bibitem [{\citenamefont {Fukuhara}\ \emph {et~al.}(2013)\citenamefont
  {Fukuhara}, \citenamefont {Schausz}, \citenamefont {Endres}, \citenamefont
  {Hild}, \citenamefont {Cheneau}, \citenamefont {Bloch},\ and\ \citenamefont
  {Gross}}]{Fukuhara2013}%
  \BibitemOpen
  \bibfield  {author} {\bibinfo {author} {\bibfnamefont {T.}~\bibnamefont
  {Fukuhara}}, \bibinfo {author} {\bibfnamefont {P.}~\bibnamefont {Schausz}},
  \bibinfo {author} {\bibfnamefont {M.}~\bibnamefont {Endres}}, \bibinfo
  {author} {\bibfnamefont {S.}~\bibnamefont {Hild}}, \bibinfo {author}
  {\bibfnamefont {M.}~\bibnamefont {Cheneau}}, \bibinfo {author} {\bibfnamefont
  {I.}~\bibnamefont {Bloch}}, \ and\ \bibinfo {author} {\bibfnamefont
  {C.}~\bibnamefont {Gross}},\ }\href {http://dx.doi.org/10.1038/nature12541}
  {\bibfield  {journal} {\bibinfo  {journal} {Nature}\ }\textbf {\bibinfo
  {volume} {502}},\ \bibinfo {pages} {76} (\bibinfo {year} {2013})},\ \bibinfo
  {note} {letter}\BibitemShut {NoStop}%
\bibitem [{\citenamefont {Ganahl}\ \emph {et~al.}(2012)\citenamefont {Ganahl},
  \citenamefont {Rabel}, \citenamefont {Essler},\ and\ \citenamefont
  {Evertz}}]{Ganahl2012}%
  \BibitemOpen
  \bibfield  {author} {\bibinfo {author} {\bibfnamefont {M.}~\bibnamefont
  {Ganahl}}, \bibinfo {author} {\bibfnamefont {E.}~\bibnamefont {Rabel}},
  \bibinfo {author} {\bibfnamefont {F.~H.~L.}\ \bibnamefont {Essler}}, \ and\
  \bibinfo {author} {\bibfnamefont {H.~G.}\ \bibnamefont {Evertz}},\ }\href
  {\doibase 10.1103/PhysRevLett.108.077206} {\bibfield  {journal} {\bibinfo
  {journal} {Phys. Rev. Lett.}\ }\textbf {\bibinfo {volume} {108}},\ \bibinfo
  {pages} {077206} (\bibinfo {year} {2012})}\BibitemShut {NoStop}%
\bibitem [{\citenamefont {Meinecke}\ \emph {et~al.}(2013)\citenamefont
  {Meinecke}, \citenamefont {Poulios}, \citenamefont {Politi}, \citenamefont
  {Matthews}, \citenamefont {Peruzzo}, \citenamefont {Ismail}, \citenamefont
  {W\"orhoff}, \citenamefont {O'Brien},\ and\ \citenamefont
  {Thompson}}]{Meinecke2013}%
  \BibitemOpen
  \bibfield  {author} {\bibinfo {author} {\bibfnamefont {J.~D.~A.}\
  \bibnamefont {Meinecke}}, \bibinfo {author} {\bibfnamefont {K.}~\bibnamefont
  {Poulios}}, \bibinfo {author} {\bibfnamefont {A.}~\bibnamefont {Politi}},
  \bibinfo {author} {\bibfnamefont {J.~C.~F.}\ \bibnamefont {Matthews}},
  \bibinfo {author} {\bibfnamefont {A.}~\bibnamefont {Peruzzo}}, \bibinfo
  {author} {\bibfnamefont {N.}~\bibnamefont {Ismail}}, \bibinfo {author}
  {\bibfnamefont {K.}~\bibnamefont {W\"orhoff}}, \bibinfo {author}
  {\bibfnamefont {J.~L.}\ \bibnamefont {O'Brien}}, \ and\ \bibinfo {author}
  {\bibfnamefont {M.~G.}\ \bibnamefont {Thompson}},\ }\href {\doibase
  10.1103/PhysRevA.88.012308} {\bibfield  {journal} {\bibinfo  {journal} {Phys.
  Rev. A}\ }\textbf {\bibinfo {volume} {88}},\ \bibinfo {pages} {012308}
  (\bibinfo {year} {2013})}\BibitemShut {NoStop}%
\bibitem [{\citenamefont {Liu}\ and\ \citenamefont {Andrei}(2014)}]{Liu2014}%
  \BibitemOpen
  \bibfield  {author} {\bibinfo {author} {\bibfnamefont {W.}~\bibnamefont
  {Liu}}\ and\ \bibinfo {author} {\bibfnamefont {N.}~\bibnamefont {Andrei}},\
  }\href {\doibase 10.1103/PhysRevLett.112.257204} {\bibfield  {journal}
  {\bibinfo  {journal} {Phys. Rev. Lett.}\ }\textbf {\bibinfo {volume} {112}},\
  \bibinfo {pages} {257204} (\bibinfo {year} {2014})}\BibitemShut {NoStop}%
\bibitem [{\citenamefont {Preiss}\ \emph {et~al.}(2015)\citenamefont {Preiss},
  \citenamefont {Ma}, \citenamefont {Tai}, \citenamefont {Lukin}, \citenamefont
  {Rispoli}, \citenamefont {Zupancic}, \citenamefont {Lahini}, \citenamefont
  {Islam},\ and\ \citenamefont {Greiner}}]{Preiss2015}%
  \BibitemOpen
  \bibfield  {author} {\bibinfo {author} {\bibfnamefont {P.~M.}\ \bibnamefont
  {Preiss}}, \bibinfo {author} {\bibfnamefont {R.}~\bibnamefont {Ma}}, \bibinfo
  {author} {\bibfnamefont {M.~E.}\ \bibnamefont {Tai}}, \bibinfo {author}
  {\bibfnamefont {A.}~\bibnamefont {Lukin}}, \bibinfo {author} {\bibfnamefont
  {M.}~\bibnamefont {Rispoli}}, \bibinfo {author} {\bibfnamefont
  {P.}~\bibnamefont {Zupancic}}, \bibinfo {author} {\bibfnamefont
  {Y.}~\bibnamefont {Lahini}}, \bibinfo {author} {\bibfnamefont
  {R.}~\bibnamefont {Islam}}, \ and\ \bibinfo {author} {\bibfnamefont
  {M.}~\bibnamefont {Greiner}},\ }\href {\doibase 10.1126/science.1260364}
  {\bibfield  {journal} {\bibinfo  {journal} {Science}\ }\textbf {\bibinfo
  {volume} {347}},\ \bibinfo {pages} {1229} (\bibinfo {year} {2015})},\ \Eprint
  {http://arxiv.org/abs/http://science.sciencemag.org/content/347/6227/1229.full.pdf}
  {http://science.sciencemag.org/content/347/6227/1229.full.pdf} \BibitemShut
  {NoStop}%
\bibitem [{\citenamefont {Wang}\ \emph {et~al.}(2015)\citenamefont {Wang},
  \citenamefont {Liu}, \citenamefont {Chen},\ and\ \citenamefont
  {Zhang}}]{Wang2015}%
  \BibitemOpen
  \bibfield  {author} {\bibinfo {author} {\bibfnamefont {L.}~\bibnamefont
  {Wang}}, \bibinfo {author} {\bibfnamefont {N.}~\bibnamefont {Liu}}, \bibinfo
  {author} {\bibfnamefont {S.}~\bibnamefont {Chen}}, \ and\ \bibinfo {author}
  {\bibfnamefont {Y.}~\bibnamefont {Zhang}},\ }\href {\doibase
  10.1103/PhysRevA.92.053606} {\bibfield  {journal} {\bibinfo  {journal} {Phys.
  Rev. A}\ }\textbf {\bibinfo {volume} {92}},\ \bibinfo {pages} {053606}
  (\bibinfo {year} {2015})}\BibitemShut {NoStop}%
\bibitem [{\citenamefont {Qin}\ \emph {et~al.}(2014)\citenamefont {Qin},
  \citenamefont {Ke}, \citenamefont {Guan}, \citenamefont {Li}, \citenamefont
  {Andrei},\ and\ \citenamefont {Lee}}]{Qin2014}%
  \BibitemOpen
  \bibfield  {author} {\bibinfo {author} {\bibfnamefont {X.}~\bibnamefont
  {Qin}}, \bibinfo {author} {\bibfnamefont {Y.}~\bibnamefont {Ke}}, \bibinfo
  {author} {\bibfnamefont {X.}~\bibnamefont {Guan}}, \bibinfo {author}
  {\bibfnamefont {Z.}~\bibnamefont {Li}}, \bibinfo {author} {\bibfnamefont
  {N.}~\bibnamefont {Andrei}}, \ and\ \bibinfo {author} {\bibfnamefont
  {C.}~\bibnamefont {Lee}},\ }\href {\doibase 10.1103/PhysRevA.90.062301}
  {\bibfield  {journal} {\bibinfo  {journal} {Phys. Rev. A}\ }\textbf {\bibinfo
  {volume} {90}},\ \bibinfo {pages} {062301} (\bibinfo {year}
  {2014})}\BibitemShut {NoStop}%
\bibitem [{\citenamefont {Benedetti}\ \emph {et~al.}(2016)\citenamefont
  {Benedetti}, \citenamefont {Buscemi}, \citenamefont {Bordone},\ and\
  \citenamefont {Paris}}]{Benedetti2016}%
  \BibitemOpen
  \bibfield  {author} {\bibinfo {author} {\bibfnamefont {C.}~\bibnamefont
  {Benedetti}}, \bibinfo {author} {\bibfnamefont {F.}~\bibnamefont {Buscemi}},
  \bibinfo {author} {\bibfnamefont {P.}~\bibnamefont {Bordone}}, \ and\
  \bibinfo {author} {\bibfnamefont {M.~G.~A.}\ \bibnamefont {Paris}},\ }\href
  {\doibase 10.1103/PhysRevA.93.042313} {\bibfield  {journal} {\bibinfo
  {journal} {Phys. Rev. A}\ }\textbf {\bibinfo {volume} {93}},\ \bibinfo
  {pages} {042313} (\bibinfo {year} {2016})}\BibitemShut {NoStop}%
\bibitem [{\citenamefont {Siloi}\ \emph {et~al.}(2017)\citenamefont {Siloi},
  \citenamefont {Benedetti}, \citenamefont {Piccinini}, \citenamefont {Piilo},
  \citenamefont {Maniscalco}, \citenamefont {Paris},\ and\ \citenamefont
  {Bordone}}]{Siloi2017}%
  \BibitemOpen
  \bibfield  {author} {\bibinfo {author} {\bibfnamefont {I.}~\bibnamefont
  {Siloi}}, \bibinfo {author} {\bibfnamefont {C.}~\bibnamefont {Benedetti}},
  \bibinfo {author} {\bibfnamefont {E.}~\bibnamefont {Piccinini}}, \bibinfo
  {author} {\bibfnamefont {J.}~\bibnamefont {Piilo}}, \bibinfo {author}
  {\bibfnamefont {S.}~\bibnamefont {Maniscalco}}, \bibinfo {author}
  {\bibfnamefont {M.~G.~A.}\ \bibnamefont {Paris}}, \ and\ \bibinfo {author}
  {\bibfnamefont {P.}~\bibnamefont {Bordone}},\ }\href {\doibase
  10.1103/PhysRevA.95.022106} {\bibfield  {journal} {\bibinfo  {journal} {Phys.
  Rev. A}\ }\textbf {\bibinfo {volume} {95}},\ \bibinfo {pages} {022106}
  (\bibinfo {year} {2017})}\BibitemShut {NoStop}%
\bibitem [{\citenamefont {Piccinini}\ \emph {et~al.}(2017)\citenamefont
  {Piccinini}, \citenamefont {Benedetti}, \citenamefont {Siloi}, \citenamefont
  {Paris},\ and\ \citenamefont {Bordone}}]{Piccinini2017}%
  \BibitemOpen
  \bibfield  {author} {\bibinfo {author} {\bibfnamefont {E.}~\bibnamefont
  {Piccinini}}, \bibinfo {author} {\bibfnamefont {C.}~\bibnamefont
  {Benedetti}}, \bibinfo {author} {\bibfnamefont {I.}~\bibnamefont {Siloi}},
  \bibinfo {author} {\bibfnamefont {M.~G.}\ \bibnamefont {Paris}}, \ and\
  \bibinfo {author} {\bibfnamefont {P.}~\bibnamefont {Bordone}},\ }\href
  {\doibase https://doi.org/10.1016/j.cpc.2017.02.014} {\bibfield  {journal}
  {\bibinfo  {journal} {Computer Physics Communications}\ }\textbf {\bibinfo
  {volume} {215}},\ \bibinfo {pages} {235 } (\bibinfo {year}
  {2017})}\BibitemShut {NoStop}%
\bibitem [{\citenamefont {Jaksch}\ \emph {et~al.}(1998)\citenamefont {Jaksch},
  \citenamefont {Bruder}, \citenamefont {Cirac}, \citenamefont {Gardiner},\
  and\ \citenamefont {Zoller}}]{Jaksch1998}%
  \BibitemOpen
  \bibfield  {author} {\bibinfo {author} {\bibfnamefont {D.}~\bibnamefont
  {Jaksch}}, \bibinfo {author} {\bibfnamefont {C.}~\bibnamefont {Bruder}},
  \bibinfo {author} {\bibfnamefont {J.~I.}\ \bibnamefont {Cirac}}, \bibinfo
  {author} {\bibfnamefont {C.~W.}\ \bibnamefont {Gardiner}}, \ and\ \bibinfo
  {author} {\bibfnamefont {P.}~\bibnamefont {Zoller}},\ }\href {\doibase
  10.1103/PhysRevLett.81.3108} {\bibfield  {journal} {\bibinfo  {journal}
  {Phys. Rev. Lett.}\ }\textbf {\bibinfo {volume} {81}},\ \bibinfo {pages}
  {3108} (\bibinfo {year} {1998})}\BibitemShut {NoStop}%
\bibitem [{\citenamefont {Zwerger}(2003)}]{Zwerger2003}%
  \BibitemOpen
  \bibfield  {author} {\bibinfo {author} {\bibfnamefont {W.}~\bibnamefont
  {Zwerger}},\ }\href {http://stacks.iop.org/1464-4266/5/i=2/a=352} {\bibfield
  {journal} {\bibinfo  {journal} {Journal of Optics B: Quantum and
  Semiclassical Optics}\ }\textbf {\bibinfo {volume} {5}},\ \bibinfo {pages}
  {S9} (\bibinfo {year} {2003})}\BibitemShut {NoStop}%
\bibitem [{\citenamefont {Dutta}\ \emph {et~al.}(2015)\citenamefont {Dutta},
  \citenamefont {Gajda}, \citenamefont {Hauke}, \citenamefont {Lewenstein},
  \citenamefont {Lühmann}, \citenamefont {Malomed}, \citenamefont
  {Sowiński},\ and\ \citenamefont {Zakrzewski}}]{Dutta2015}%
  \BibitemOpen
  \bibfield  {author} {\bibinfo {author} {\bibfnamefont {O.}~\bibnamefont
  {Dutta}}, \bibinfo {author} {\bibfnamefont {M.}~\bibnamefont {Gajda}},
  \bibinfo {author} {\bibfnamefont {P.}~\bibnamefont {Hauke}}, \bibinfo
  {author} {\bibfnamefont {M.}~\bibnamefont {Lewenstein}}, \bibinfo {author}
  {\bibfnamefont {D.-S.}\ \bibnamefont {Lühmann}}, \bibinfo {author}
  {\bibfnamefont {B.~A.}\ \bibnamefont {Malomed}}, \bibinfo {author}
  {\bibfnamefont {T.}~\bibnamefont {Sowiński}}, \ and\ \bibinfo {author}
  {\bibfnamefont {J.}~\bibnamefont {Zakrzewski}},\ }\href
  {http://stacks.iop.org/0034-4885/78/i=6/a=066001} {\bibfield  {journal}
  {\bibinfo  {journal} {Reports on Progress in Physics}\ }\textbf {\bibinfo
  {volume} {78}},\ \bibinfo {pages} {066001} (\bibinfo {year}
  {2015})}\BibitemShut {NoStop}%
\bibitem [{\citenamefont {Glück}\ \emph {et~al.}(2002)\citenamefont {Glück},
  \citenamefont {Kolovsky},\ and\ \citenamefont {Korsch}}]{Gluck2002}%
  \BibitemOpen
  \bibfield  {author} {\bibinfo {author} {\bibfnamefont {M.}~\bibnamefont
  {Glück}}, \bibinfo {author} {\bibfnamefont {A.~R.}\ \bibnamefont
  {Kolovsky}}, \ and\ \bibinfo {author} {\bibfnamefont {H.~J.}\ \bibnamefont
  {Korsch}},\ }\href {\doibase http://doi.org/10.1016/S0370-1573(02)00142-4}
  {\bibfield  {journal} {\bibinfo  {journal} {Physics Reports}\ }\textbf
  {\bibinfo {volume} {366}},\ \bibinfo {pages} {103 } (\bibinfo {year}
  {2002})}\BibitemShut {NoStop}%
\bibitem [{\citenamefont {Sachdev}\ \emph {et~al.}(2002)\citenamefont
  {Sachdev}, \citenamefont {Sengupta},\ and\ \citenamefont
  {Girvin}}]{Sachdev2002}%
  \BibitemOpen
  \bibfield  {author} {\bibinfo {author} {\bibfnamefont {S.}~\bibnamefont
  {Sachdev}}, \bibinfo {author} {\bibfnamefont {K.}~\bibnamefont {Sengupta}}, \
  and\ \bibinfo {author} {\bibfnamefont {S.~M.}\ \bibnamefont {Girvin}},\
  }\href {\doibase 10.1103/PhysRevB.66.075128} {\bibfield  {journal} {\bibinfo
  {journal} {Phys. Rev. B}\ }\textbf {\bibinfo {volume} {66}},\ \bibinfo
  {pages} {075128} (\bibinfo {year} {2002})}\BibitemShut {NoStop}%
\bibitem [{\citenamefont {Kolovsky}(2003)}]{Kolovsky2003}%
  \BibitemOpen
  \bibfield  {author} {\bibinfo {author} {\bibfnamefont {A.~R.}\ \bibnamefont
  {Kolovsky}},\ }\href {\doibase 10.1103/PhysRevLett.90.213002} {\bibfield
  {journal} {\bibinfo  {journal} {Phys. Rev. Lett.}\ }\textbf {\bibinfo
  {volume} {90}},\ \bibinfo {pages} {213002} (\bibinfo {year}
  {2003})}\BibitemShut {NoStop}%
\bibitem [{\citenamefont {Buchleitner}\ and\ \citenamefont
  {Kolovsky}(2003)}]{Buchleitner2003}%
  \BibitemOpen
  \bibfield  {author} {\bibinfo {author} {\bibfnamefont {A.}~\bibnamefont
  {Buchleitner}}\ and\ \bibinfo {author} {\bibfnamefont {A.~R.}\ \bibnamefont
  {Kolovsky}},\ }\href {\doibase 10.1103/PhysRevLett.91.253002} {\bibfield
  {journal} {\bibinfo  {journal} {Phys. Rev. Lett.}\ }\textbf {\bibinfo
  {volume} {91}},\ \bibinfo {pages} {253002} (\bibinfo {year}
  {2003})}\BibitemShut {NoStop}%
\bibitem [{\citenamefont {Kolovsky}(2004)}]{Kolovsky04}%
  \BibitemOpen
  \bibfield  {author} {\bibinfo {author} {\bibfnamefont {A.~R.}\ \bibnamefont
  {Kolovsky}},\ }\href {\doibase 10.1103/PhysRevA.70.015604} {\bibfield
  {journal} {\bibinfo  {journal} {Phys. Rev. A}\ }\textbf {\bibinfo {volume}
  {70}},\ \bibinfo {pages} {015604} (\bibinfo {year} {2004})}\BibitemShut
  {NoStop}%
\bibitem [{\citenamefont {Kolovsky}\ \emph {et~al.}(2010)\citenamefont
  {Kolovsky}, \citenamefont {G\'omez},\ and\ \citenamefont
  {Korsch}}]{Kolovsky2010}%
  \BibitemOpen
  \bibfield  {author} {\bibinfo {author} {\bibfnamefont {A.~R.}\ \bibnamefont
  {Kolovsky}}, \bibinfo {author} {\bibfnamefont {E.~A.}\ \bibnamefont
  {G\'omez}}, \ and\ \bibinfo {author} {\bibfnamefont {H.~J.}\ \bibnamefont
  {Korsch}},\ }\href {\doibase 10.1103/PhysRevA.81.025603} {\bibfield
  {journal} {\bibinfo  {journal} {Phys. Rev. A}\ }\textbf {\bibinfo {volume}
  {81}},\ \bibinfo {pages} {025603} (\bibinfo {year} {2010})}\BibitemShut
  {NoStop}%
\bibitem [{\citenamefont {D\'{\i}az}\ \emph {et~al.}(2013)\citenamefont
  {D\'{\i}az}, \citenamefont {Garc\'{\i}a~Mena}, \citenamefont {Asakura},\ and\
  \citenamefont {Gaul}}]{Diaz2013}%
  \BibitemOpen
  \bibfield  {author} {\bibinfo {author} {\bibfnamefont {E.}~\bibnamefont
  {D\'{\i}az}}, \bibinfo {author} {\bibfnamefont {A.}~\bibnamefont
  {Garc\'{\i}a~Mena}}, \bibinfo {author} {\bibfnamefont {K.}~\bibnamefont
  {Asakura}}, \ and\ \bibinfo {author} {\bibfnamefont {C.}~\bibnamefont
  {Gaul}},\ }\href {\doibase 10.1103/PhysRevA.87.015601} {\bibfield  {journal}
  {\bibinfo  {journal} {Phys. Rev. A}\ }\textbf {\bibinfo {volume} {87}},\
  \bibinfo {pages} {015601} (\bibinfo {year} {2013})}\BibitemShut {NoStop}%
\bibitem [{\citenamefont {Mandt}(2014)}]{Mandt2014}%
  \BibitemOpen
  \bibfield  {author} {\bibinfo {author} {\bibfnamefont {S.}~\bibnamefont
  {Mandt}},\ }\href {\doibase 10.1103/PhysRevA.90.053624} {\bibfield  {journal}
  {\bibinfo  {journal} {Phys. Rev. A}\ }\textbf {\bibinfo {volume} {90}},\
  \bibinfo {pages} {053624} (\bibinfo {year} {2014})}\BibitemShut {NoStop}%
\bibitem [{\citenamefont {Meinert}\ \emph {et~al.}(2014)\citenamefont
  {Meinert}, \citenamefont {Mark}, \citenamefont {Kirilov}, \citenamefont
  {Lauber}, \citenamefont {Weinmann}, \citenamefont {Gr\"obner},\ and\
  \citenamefont {N\"agerl}}]{Meinert2014}%
  \BibitemOpen
  \bibfield  {author} {\bibinfo {author} {\bibfnamefont {F.}~\bibnamefont
  {Meinert}}, \bibinfo {author} {\bibfnamefont {M.~J.}\ \bibnamefont {Mark}},
  \bibinfo {author} {\bibfnamefont {E.}~\bibnamefont {Kirilov}}, \bibinfo
  {author} {\bibfnamefont {K.}~\bibnamefont {Lauber}}, \bibinfo {author}
  {\bibfnamefont {P.}~\bibnamefont {Weinmann}}, \bibinfo {author}
  {\bibfnamefont {M.}~\bibnamefont {Gr\"obner}}, \ and\ \bibinfo {author}
  {\bibfnamefont {H.-C.}\ \bibnamefont {N\"agerl}},\ }\href {\doibase
  10.1103/PhysRevLett.112.193003} {\bibfield  {journal} {\bibinfo  {journal}
  {Phys. Rev. Lett.}\ }\textbf {\bibinfo {volume} {112}},\ \bibinfo {pages}
  {193003} (\bibinfo {year} {2014})}\BibitemShut {NoStop}%
\bibitem [{\citenamefont {Huse}\ \emph {et~al.}(2014)\citenamefont {Huse},
  \citenamefont {Nandkishore},\ and\ \citenamefont {Oganesyan}}]{Huse14}%
  \BibitemOpen
  \bibfield  {author} {\bibinfo {author} {\bibfnamefont {D.~A.}\ \bibnamefont
  {Huse}}, \bibinfo {author} {\bibfnamefont {R.}~\bibnamefont {Nandkishore}}, \
  and\ \bibinfo {author} {\bibfnamefont {V.}~\bibnamefont {Oganesyan}},\ }\href
  {\doibase 10.1103/PhysRevB.90.174202} {\bibfield  {journal} {\bibinfo
  {journal} {Phys. Rev. B}\ }\textbf {\bibinfo {volume} {90}},\ \bibinfo
  {pages} {174202} (\bibinfo {year} {2014})}\BibitemShut {NoStop}%
\bibitem [{\citenamefont {Nandkishore}\ and\ \citenamefont
  {Huse}(2015)}]{Rahul15}%
  \BibitemOpen
  \bibfield  {author} {\bibinfo {author} {\bibfnamefont {R.}~\bibnamefont
  {Nandkishore}}\ and\ \bibinfo {author} {\bibfnamefont {D.~A.}\ \bibnamefont
  {Huse}},\ }\href@noop {} {\bibfield  {journal} {\bibinfo  {journal} {Ann.
  Rev. Cond. Mat. Phys.}\ }\textbf {\bibinfo {volume} {6}},\ \bibinfo {pages}
  {15} (\bibinfo {year} {2015})}\BibitemShut {NoStop}%
\bibitem [{\citenamefont {Hartmann}\ \emph {et~al.}(2004)\citenamefont
  {Hartmann}, \citenamefont {Keck}, \citenamefont {Korsch},\ and\ \citenamefont
  {Mossmann}}]{Hartmann2004}%
  \BibitemOpen
  \bibfield  {author} {\bibinfo {author} {\bibfnamefont {T.}~\bibnamefont
  {Hartmann}}, \bibinfo {author} {\bibfnamefont {F.}~\bibnamefont {Keck}},
  \bibinfo {author} {\bibfnamefont {H.~J.}\ \bibnamefont {Korsch}}, \ and\
  \bibinfo {author} {\bibfnamefont {S.}~\bibnamefont {Mossmann}},\ }\href
  {http://stacks.iop.org/1367-2630/6/i=1/a=002} {\bibfield  {journal} {\bibinfo
   {journal} {New Journal of Physics}\ }\textbf {\bibinfo {volume} {6}},\
  \bibinfo {pages} {2} (\bibinfo {year} {2004})}\BibitemShut {NoStop}%
\bibitem [{\citenamefont {Dias}\ \emph {et~al.}(2007)\citenamefont {Dias},
  \citenamefont {Nascimento}, \citenamefont {Lyra},\ and\ \citenamefont
  {de~Moura}}]{Dias07}%
  \BibitemOpen
  \bibfield  {author} {\bibinfo {author} {\bibfnamefont {W.~S.}\ \bibnamefont
  {Dias}}, \bibinfo {author} {\bibfnamefont {E.~M.}\ \bibnamefont
  {Nascimento}}, \bibinfo {author} {\bibfnamefont {M.~L.}\ \bibnamefont
  {Lyra}}, \ and\ \bibinfo {author} {\bibfnamefont {F.~A. B.~F.}\ \bibnamefont
  {de~Moura}},\ }\href {\doibase 10.1103/PhysRevB.76.155124} {\bibfield
  {journal} {\bibinfo  {journal} {Phys. Rev. B}\ }\textbf {\bibinfo {volume}
  {76}},\ \bibinfo {pages} {155124} (\bibinfo {year} {2007})}\BibitemShut
  {NoStop}%
\bibitem [{\citenamefont {Khomeriki}\ \emph {et~al.}(2010)\citenamefont
  {Khomeriki}, \citenamefont {Krimer}, \citenamefont {Haque},\ and\
  \citenamefont {Flach}}]{Khomeriki10}%
  \BibitemOpen
  \bibfield  {author} {\bibinfo {author} {\bibfnamefont {R.}~\bibnamefont
  {Khomeriki}}, \bibinfo {author} {\bibfnamefont {D.~O.}\ \bibnamefont
  {Krimer}}, \bibinfo {author} {\bibfnamefont {M.}~\bibnamefont {Haque}}, \
  and\ \bibinfo {author} {\bibfnamefont {S.}~\bibnamefont {Flach}},\ }\href
  {\doibase 10.1103/PhysRevA.81.065601} {\bibfield  {journal} {\bibinfo
  {journal} {Phys. Rev. A}\ }\textbf {\bibinfo {volume} {81}},\ \bibinfo
  {pages} {065601} (\bibinfo {year} {2010})}\BibitemShut {NoStop}%
\bibitem [{\citenamefont {Ben~Dahan}\ \emph {et~al.}(1996)\citenamefont
  {Ben~Dahan}, \citenamefont {Peik}, \citenamefont {Reichel}, \citenamefont
  {Castin},\ and\ \citenamefont {Salomon}}]{Dahan96}%
  \BibitemOpen
  \bibfield  {author} {\bibinfo {author} {\bibfnamefont {M.}~\bibnamefont
  {Ben~Dahan}}, \bibinfo {author} {\bibfnamefont {E.}~\bibnamefont {Peik}},
  \bibinfo {author} {\bibfnamefont {J.}~\bibnamefont {Reichel}}, \bibinfo
  {author} {\bibfnamefont {Y.}~\bibnamefont {Castin}}, \ and\ \bibinfo {author}
  {\bibfnamefont {C.}~\bibnamefont {Salomon}},\ }\href {\doibase
  10.1103/PhysRevLett.76.4508} {\bibfield  {journal} {\bibinfo  {journal}
  {Phys. Rev. Lett.}\ }\textbf {\bibinfo {volume} {76}},\ \bibinfo {pages}
  {4508} (\bibinfo {year} {1996})}\BibitemShut {NoStop}%
\bibitem [{\citenamefont {Lyssenko}\ \emph {et~al.}(1997)\citenamefont
  {Lyssenko}, \citenamefont {Valu\ifmmode~\check{s}\else \v{s}\fi{}is},
  \citenamefont {L\"oser}, \citenamefont {Hasche}, \citenamefont {Leo},
  \citenamefont {Dignam},\ and\ \citenamefont {K\"ohler}}]{Lyssenko97}%
  \BibitemOpen
  \bibfield  {author} {\bibinfo {author} {\bibfnamefont {V.~G.}\ \bibnamefont
  {Lyssenko}}, \bibinfo {author} {\bibfnamefont {G.}~\bibnamefont
  {Valu\ifmmode~\check{s}\else \v{s}\fi{}is}}, \bibinfo {author} {\bibfnamefont
  {F.}~\bibnamefont {L\"oser}}, \bibinfo {author} {\bibfnamefont
  {T.}~\bibnamefont {Hasche}}, \bibinfo {author} {\bibfnamefont
  {K.}~\bibnamefont {Leo}}, \bibinfo {author} {\bibfnamefont {M.~M.}\
  \bibnamefont {Dignam}}, \ and\ \bibinfo {author} {\bibfnamefont
  {K.}~\bibnamefont {K\"ohler}},\ }\href {\doibase 10.1103/PhysRevLett.79.301}
  {\bibfield  {journal} {\bibinfo  {journal} {Phys. Rev. Lett.}\ }\textbf
  {\bibinfo {volume} {79}},\ \bibinfo {pages} {301} (\bibinfo {year}
  {1997})}\BibitemShut {NoStop}%
\bibitem [{\citenamefont {Zenesini}\ \emph {et~al.}(2009)\citenamefont
  {Zenesini}, \citenamefont {Lignier}, \citenamefont {Tayebirad}, \citenamefont
  {Radogostowicz}, \citenamefont {Ciampini}, \citenamefont {Mannella},
  \citenamefont {Wimberger}, \citenamefont {Morsch},\ and\ \citenamefont
  {Arimondo}}]{Zanesini09}%
  \BibitemOpen
  \bibfield  {author} {\bibinfo {author} {\bibfnamefont {A.}~\bibnamefont
  {Zenesini}}, \bibinfo {author} {\bibfnamefont {H.}~\bibnamefont {Lignier}},
  \bibinfo {author} {\bibfnamefont {G.}~\bibnamefont {Tayebirad}}, \bibinfo
  {author} {\bibfnamefont {J.}~\bibnamefont {Radogostowicz}}, \bibinfo {author}
  {\bibfnamefont {D.}~\bibnamefont {Ciampini}}, \bibinfo {author}
  {\bibfnamefont {R.}~\bibnamefont {Mannella}}, \bibinfo {author}
  {\bibfnamefont {S.}~\bibnamefont {Wimberger}}, \bibinfo {author}
  {\bibfnamefont {O.}~\bibnamefont {Morsch}}, \ and\ \bibinfo {author}
  {\bibfnamefont {E.}~\bibnamefont {Arimondo}},\ }\href {\doibase
  10.1103/PhysRevLett.103.090403} {\bibfield  {journal} {\bibinfo  {journal}
  {Phys. Rev. Lett.}\ }\textbf {\bibinfo {volume} {103}},\ \bibinfo {pages}
  {090403} (\bibinfo {year} {2009})}\BibitemShut {NoStop}%
\bibitem [{\citenamefont {Poli}\ \emph {et~al.}(2011)\citenamefont {Poli},
  \citenamefont {Wang}, \citenamefont {Tarallo}, \citenamefont {Alberti},
  \citenamefont {Prevedelli},\ and\ \citenamefont {Tino}}]{Poli11}%
  \BibitemOpen
  \bibfield  {author} {\bibinfo {author} {\bibfnamefont {N.}~\bibnamefont
  {Poli}}, \bibinfo {author} {\bibfnamefont {F.-Y.}\ \bibnamefont {Wang}},
  \bibinfo {author} {\bibfnamefont {M.~G.}\ \bibnamefont {Tarallo}}, \bibinfo
  {author} {\bibfnamefont {A.}~\bibnamefont {Alberti}}, \bibinfo {author}
  {\bibfnamefont {M.}~\bibnamefont {Prevedelli}}, \ and\ \bibinfo {author}
  {\bibfnamefont {G.~M.}\ \bibnamefont {Tino}},\ }\href {\doibase
  10.1103/PhysRevLett.106.038501} {\bibfield  {journal} {\bibinfo  {journal}
  {Phys. Rev. Lett.}\ }\textbf {\bibinfo {volume} {106}},\ \bibinfo {pages}
  {038501} (\bibinfo {year} {2011})}\BibitemShut {NoStop}%
\bibitem [{\citenamefont {Schreiber}\ \emph {et~al.}(2015)\citenamefont
  {Schreiber}, \citenamefont {Hodgman}, \citenamefont {Bordia}, \citenamefont
  {L{\"u}schen}, \citenamefont {Fischer}, \citenamefont {Vosk}, \citenamefont
  {Altman}, \citenamefont {Schneider},\ and\ \citenamefont
  {Bloch}}]{Schreiber2015}%
  \BibitemOpen
  \bibfield  {author} {\bibinfo {author} {\bibfnamefont {M.}~\bibnamefont
  {Schreiber}}, \bibinfo {author} {\bibfnamefont {S.~S.}\ \bibnamefont
  {Hodgman}}, \bibinfo {author} {\bibfnamefont {P.}~\bibnamefont {Bordia}},
  \bibinfo {author} {\bibfnamefont {H.~P.}\ \bibnamefont {L{\"u}schen}},
  \bibinfo {author} {\bibfnamefont {M.~H.}\ \bibnamefont {Fischer}}, \bibinfo
  {author} {\bibfnamefont {R.}~\bibnamefont {Vosk}}, \bibinfo {author}
  {\bibfnamefont {E.}~\bibnamefont {Altman}}, \bibinfo {author} {\bibfnamefont
  {U.}~\bibnamefont {Schneider}}, \ and\ \bibinfo {author} {\bibfnamefont
  {I.}~\bibnamefont {Bloch}},\ }\href {\doibase 10.1126/science.aaa7432}
  {\bibfield  {journal} {\bibinfo  {journal} {Science}\ }\textbf {\bibinfo
  {volume} {349}},\ \bibinfo {pages} {842} (\bibinfo {year} {2015})},\ \Eprint
  {http://arxiv.org/abs/http://science.sciencemag.org/content/349/6250/842.full.pdf}
  {http://science.sciencemag.org/content/349/6250/842.full.pdf} \BibitemShut
  {NoStop}%
\bibitem [{\citenamefont {Bordia}\ \emph {et~al.}(2016)\citenamefont {Bordia},
  \citenamefont {L\"uschen}, \citenamefont {Hodgman}, \citenamefont
  {Schreiber}, \citenamefont {Bloch},\ and\ \citenamefont
  {Schneider}}]{Bordia2016}%
  \BibitemOpen
  \bibfield  {author} {\bibinfo {author} {\bibfnamefont {P.}~\bibnamefont
  {Bordia}}, \bibinfo {author} {\bibfnamefont {H.~P.}\ \bibnamefont
  {L\"uschen}}, \bibinfo {author} {\bibfnamefont {S.~S.}\ \bibnamefont
  {Hodgman}}, \bibinfo {author} {\bibfnamefont {M.}~\bibnamefont {Schreiber}},
  \bibinfo {author} {\bibfnamefont {I.}~\bibnamefont {Bloch}}, \ and\ \bibinfo
  {author} {\bibfnamefont {U.}~\bibnamefont {Schneider}},\ }\href {\doibase
  10.1103/PhysRevLett.116.140401} {\bibfield  {journal} {\bibinfo  {journal}
  {Phys. Rev. Lett.}\ }\textbf {\bibinfo {volume} {116}},\ \bibinfo {pages}
  {140401} (\bibinfo {year} {2016})}\BibitemShut {NoStop}%
\bibitem [{\citenamefont {Bordia}\ \emph {et~al.}(2017)\citenamefont {Bordia},
  \citenamefont {Luschen}, \citenamefont {Schneider}, \citenamefont {Knap},\
  and\ \citenamefont {Bloch}}]{Bordia2017}%
  \BibitemOpen
  \bibfield  {author} {\bibinfo {author} {\bibfnamefont {P.}~\bibnamefont
  {Bordia}}, \bibinfo {author} {\bibfnamefont {H.}~\bibnamefont {Luschen}},
  \bibinfo {author} {\bibfnamefont {U.}~\bibnamefont {Schneider}}, \bibinfo
  {author} {\bibfnamefont {M.}~\bibnamefont {Knap}}, \ and\ \bibinfo {author}
  {\bibfnamefont {I.}~\bibnamefont {Bloch}},\ }\href
  {http://dx.doi.org/10.1038/nphys4020} {\bibfield  {journal} {\bibinfo
  {journal} {Nat Phys}\ }\textbf {\bibinfo {volume} {advance online
  publication}} (\bibinfo {year} {2017})},\ \bibinfo {note}
  {article}\BibitemShut {NoStop}%
\bibitem [{\citenamefont {{Aubry}}\ and\ \citenamefont
  {{Andr\'e}}(1980)}]{AA80}%
  \BibitemOpen
  \bibfield  {author} {\bibinfo {author} {\bibfnamefont {S.}~\bibnamefont
  {{Aubry}}}\ and\ \bibinfo {author} {\bibfnamefont {G.}~\bibnamefont
  {{Andr\'e}}},\ }\href@noop {} {\bibfield  {journal} {\bibinfo  {journal}
  {Ann. Israel Phys. Soc}\ }\textbf {\bibinfo {volume} {3}},\ \bibinfo {pages}
  {18} (\bibinfo {year} {1980})}\BibitemShut {NoStop}%
\bibitem [{\citenamefont {M\"uller}\ and\ \citenamefont
  {Delande}(2011)}]{Mueller2009}%
  \BibitemOpen
  \bibfield  {author} {\bibinfo {author} {\bibfnamefont {C.~A.}\ \bibnamefont
  {M\"uller}}\ and\ \bibinfo {author} {\bibfnamefont {D.}~\bibnamefont
  {Delande}},\ }\enquote {\bibinfo {title} {{Disorder and interference:
  localization phenomena}},}\ in\ \href {\doibase
  DOI:10.1093/acprof:oso/9780199603657.003.0009} {\emph {\bibinfo {booktitle}
  {Lecture Notes of the Les Houches Summer School in Singapore: Ultracold Gases
  and Quantum Information}}},\ Vol.~\bibinfo {volume} {91}\ (\bibinfo
  {publisher} {Oxford Scholarship},\ \bibinfo {year} {2011})\ Chap.~\bibinfo
  {chapter} {9},\ \Eprint {http://arxiv.org/abs/1005.0915} {arXiv:1005.0915}
  \BibitemShut {NoStop}%
\bibitem [{\citenamefont {Basko}\ \emph {et~al.}(2006)\citenamefont {Basko},
  \citenamefont {Aleiner},\ and\ \citenamefont {Altschuler}}]{Basko06}%
  \BibitemOpen
  \bibfield  {author} {\bibinfo {author} {\bibfnamefont {D.}~\bibnamefont
  {Basko}}, \bibinfo {author} {\bibfnamefont {I.}~\bibnamefont {Aleiner}}, \
  and\ \bibinfo {author} {\bibfnamefont {B.}~\bibnamefont {Altschuler}},\
  }\href@noop {} {\bibfield  {journal} {\bibinfo  {journal} {Ann. Phys. (NY)}\
  }\textbf {\bibinfo {volume} {321}},\ \bibinfo {pages} {1126} (\bibinfo {year}
  {2006})}\BibitemShut {NoStop}%
\bibitem [{\citenamefont {Srednicki}(1994)}]{srednicki94}%
  \BibitemOpen
  \bibfield  {author} {\bibinfo {author} {\bibfnamefont {M.}~\bibnamefont
  {Srednicki}},\ }\href {\doibase 10.1103/PhysRevE.50.888} {\bibfield
  {journal} {\bibinfo  {journal} {Phys. Rev. E}\ }\textbf {\bibinfo {volume}
  {50}},\ \bibinfo {pages} {888} (\bibinfo {year} {1994})}\BibitemShut
  {NoStop}%
\bibitem [{\citenamefont {Serbyn}\ \emph {et~al.}(2013)\citenamefont {Serbyn},
  \citenamefont {Papi\ifmmode~\acute{c}\else \'{c}\fi{}},\ and\ \citenamefont
  {Abanin}}]{Serbyn2013}%
  \BibitemOpen
  \bibfield  {author} {\bibinfo {author} {\bibfnamefont {M.}~\bibnamefont
  {Serbyn}}, \bibinfo {author} {\bibfnamefont {Z.}~\bibnamefont
  {Papi\ifmmode~\acute{c}\else \'{c}\fi{}}}, \ and\ \bibinfo {author}
  {\bibfnamefont {D.~A.}\ \bibnamefont {Abanin}},\ }\href {\doibase
  10.1103/PhysRevLett.111.127201} {\bibfield  {journal} {\bibinfo  {journal}
  {Phys. Rev. Lett.}\ }\textbf {\bibinfo {volume} {111}},\ \bibinfo {pages}
  {127201} (\bibinfo {year} {2013})}\BibitemShut {NoStop}%
\bibitem [{\citenamefont {\ifmmode \check{Z}\else
  \v{Z}\fi{}nidari\ifmmode~\check{c}\else \v{c}\fi{}}\ \emph
  {et~al.}(2008)\citenamefont {\ifmmode \check{Z}\else
  \v{Z}\fi{}nidari\ifmmode~\check{c}\else \v{c}\fi{}}, \citenamefont {Prosen},\
  and\ \citenamefont {Prelov\ifmmode~\check{s}\else
  \v{s}\fi{}ek}}]{Znidaric2008}%
  \BibitemOpen
  \bibfield  {author} {\bibinfo {author} {\bibfnamefont {M.}~\bibnamefont
  {\ifmmode \check{Z}\else \v{Z}\fi{}nidari\ifmmode~\check{c}\else
  \v{c}\fi{}}}, \bibinfo {author} {\bibfnamefont {T.~c.~v.}\ \bibnamefont
  {Prosen}}, \ and\ \bibinfo {author} {\bibfnamefont {P.}~\bibnamefont
  {Prelov\ifmmode~\check{s}\else \v{s}\fi{}ek}},\ }\href {\doibase
  10.1103/PhysRevB.77.064426} {\bibfield  {journal} {\bibinfo  {journal} {Phys.
  Rev. B}\ }\textbf {\bibinfo {volume} {77}},\ \bibinfo {pages} {064426}
  (\bibinfo {year} {2008})}\BibitemShut {NoStop}%
\bibitem [{\citenamefont {Bardarson}\ \emph {et~al.}(2012)\citenamefont
  {Bardarson}, \citenamefont {Pollmann},\ and\ \citenamefont
  {Moore}}]{Bardarson2012}%
  \BibitemOpen
  \bibfield  {author} {\bibinfo {author} {\bibfnamefont {J.~H.}\ \bibnamefont
  {Bardarson}}, \bibinfo {author} {\bibfnamefont {F.}~\bibnamefont {Pollmann}},
  \ and\ \bibinfo {author} {\bibfnamefont {J.~E.}\ \bibnamefont {Moore}},\
  }\href {\doibase 10.1103/PhysRevLett.109.017202} {\bibfield  {journal}
  {\bibinfo  {journal} {Phys. Rev. Lett.}\ }\textbf {\bibinfo {volume} {109}},\
  \bibinfo {pages} {017202} (\bibinfo {year} {2012})}\BibitemShut {NoStop}%
\bibitem [{\citenamefont {{Diez}}\ \emph {et~al.}(1996)\citenamefont {{Diez}},
  \citenamefont {{Dominguez-Adame}},\ and\ \citenamefont {{Sanchez}}}]{Diez96}%
  \BibitemOpen
  \bibfield  {author} {\bibinfo {author} {\bibfnamefont {E.}~\bibnamefont
  {{Diez}}}, \bibinfo {author} {\bibfnamefont {F.}~\bibnamefont
  {{Dominguez-Adame}}}, \ and\ \bibinfo {author} {\bibfnamefont
  {A.}~\bibnamefont {{Sanchez}}},\ }\href@noop {} {\bibfield  {journal}
  {\bibinfo  {journal} {eprint arXiv:cond-mat/9610204}\ } (\bibinfo {year}
  {1996})},\ \Eprint {http://arxiv.org/abs/cond-mat/9610204} {cond-mat/9610204}
  \BibitemShut {NoStop}%
\bibitem [{\citenamefont {Schulte}\ \emph {et~al.}(2008)\citenamefont
  {Schulte}, \citenamefont {Drenkelforth}, \citenamefont {B\"uning},
  \citenamefont {Ertmer}, \citenamefont {Arlt}, \citenamefont {Lewenstein},\
  and\ \citenamefont {Santos}}]{Schulte2008}%
  \BibitemOpen
  \bibfield  {author} {\bibinfo {author} {\bibfnamefont {T.}~\bibnamefont
  {Schulte}}, \bibinfo {author} {\bibfnamefont {S.}~\bibnamefont
  {Drenkelforth}}, \bibinfo {author} {\bibfnamefont {G.~K.}\ \bibnamefont
  {B\"uning}}, \bibinfo {author} {\bibfnamefont {W.}~\bibnamefont {Ertmer}},
  \bibinfo {author} {\bibfnamefont {J.}~\bibnamefont {Arlt}}, \bibinfo {author}
  {\bibfnamefont {M.}~\bibnamefont {Lewenstein}}, \ and\ \bibinfo {author}
  {\bibfnamefont {L.}~\bibnamefont {Santos}},\ }\href {\doibase
  10.1103/PhysRevA.77.023610} {\bibfield  {journal} {\bibinfo  {journal} {Phys.
  Rev. A}\ }\textbf {\bibinfo {volume} {77}},\ \bibinfo {pages} {023610}
  (\bibinfo {year} {2008})}\BibitemShut {NoStop}%
\end{thebibliography}
%

\end{document}